\documentclass[preprint,3p,latin,12pt,authoryear]{elsarticle}

\usepackage{amssymb}
\usepackage{color}
\usepackage{xcolor}
\usepackage[english]{babel}
\usepackage{amsthm}
\theoremstyle{definition}
\newtheorem{remark}{Remark}
\usepackage{geometry}
\usepackage{siunitx}
\usepackage{hyperref}	 
\hypersetup{colorlinks=true,
	linkcolor=blue,%
	citecolor=blue}

\usepackage{algorithm}
\usepackage{algpseudocode}
\makeatletter  
\newif\if@restonecol  
\makeatother

\usepackage{graphicx}
\graphicspath{{./Figures/}}
\usepackage[normalem]{ulem}

\newcommand\mdl{\bgroup\markoverwith{\textcolor{red}{\rule[1.5ex]{2pt}{0.8pt}}}\ULon}

\usepackage[mathlines]{lineno}
\usepackage{etoolbox} 

\usepackage{caption}
\usepackage{subcaption}
\usepackage{booktabs}
\usepackage{multirow}
\usepackage[flushleft]{threeparttable}
\usepackage{bm}
\usepackage{amsmath}
\usepackage{cases}
\allowdisplaybreaks[4]
\usepackage{enumitem}
\usepackage{lscape}
\usepackage{rotfloat}
\usepackage{pdflscape}
\usepackage{makecell}
\usepackage{orcidlink}

\usepackage{tikz}
\journal{arXiv}

\begin{document}
\sloppy

\begin{frontmatter}

\title{A Virtual Heat Flux Method for Simple and Accurate Neumann Thermal Boundary Imposition in the Material Point Method} 

\author[1,2]{Jidu YU\corref{*}}
\ead{jyubu@connect.ust.hk}

\author[1]{Jidong ZHAO\corref{*}}
\ead{jzhao@ust.hk}

\cortext[*]{Corresponding author}

\affiliation[1]{organization={Department of Civil and Environmental Engineering, Hong Kong University of Science and Technology},
             state={Hong Kong},
             country={China}}             

\affiliation[2]{organization={Department of Civil and Environmental Engineering, University of California, Berkeley},
             country={USA}}  

\begin{abstract}
In the Material Point Method (MPM), accurately imposing Neumann-type thermal boundary conditions, particularly convective heat flux boundaries, remains a significant challenge due to the inherent nonconformity between complex evolving material boundaries and the fixed background grid. This paper introduces a novel Virtual Heat Flux Method (VHFM) to overcome this limitation. The core idea is to construct a virtual flux field on an auxiliary domain surrounding the physical boundary, which exactly satisfies the prescribed boundary condition. This transforms the surface integral in the weak form into an equivalent, and easily computed, volumetric integral. Consequently, VHFM eliminates the need for explicit boundary tracking, specialized boundary particles, or complex surface reconstruction. A unified formulation is presented, demonstrating the method's straightforward extension to general scalar, vector, and tensor Neumann conditions. The accuracy, robustness, and convergence of VHFM are rigorously validated through a series of numerical benchmarks, including 1D transient analysis, 2D and 3D curved boundaries, and problems with large rotations and complex moving geometries. The results show that VHFM achieves accuracy comparable to conforming node-based imposition and significantly outperforms conventional particle-based approaches. Its simplicity, computational efficiency, and robustness make it an attractive solution for integrating accurate thermal boundary conditions into thermo-mechanical and other multiphysics MPM frameworks.
\end{abstract}

\end{frontmatter}


\section{Introduction}
The Material Point Method (MPM) is a hybrid mesh–particle computational framework that combines the strengths of Lagrangian material points with Eulerian background grids \citep{sulsky1994, sulsky1995, bardenhagen2004}. Over the past several decades, MPM has undergone rapid development and has become a widely adopted tool in computational mechanics. Compared with the traditional mesh‑based Finite Element Method (FEM), MPM offers substantial advantages for simulating large deformations. Its central idea is to store material properties, such as mass, velocity, stress, and strain, on material points while employing a background grid for mapping and updating field variables. This hybrid strategy effectively mitigates numerical issues associated with mesh distortion. In addition, MPM exhibits strong versatility in modeling complex physical processes, including multiphase flows \citep{chen2018, chandra2024stabilized, kularathna2021a}, fracture and failure \citep{liang2022shear, liang2021extended}, and contact interactions \citep{liang2024mortar}. Owing to these capabilities, MPM has been successfully applied across a broad spectrum of disciplines, such as geohazard analysis (e.g., landslides, debris flows, and avalanches) \citep{soga2016, Gaume2017snow, Gaume2018dynamic, Gaume2019investigating, Li2021three, Alonso2021}, soil-fluid–structure interaction \citep{cheng2026stabilized, liang2021, liang2021a}, biomechanics \citep{guilkey2006computational, li2021novel}, and computer graphics \citep{Stomakhin2013, Stomakhin2014, Liang2019, Su2021}.

In recent years, the classical MPM framework \citep{sulsky1994} has been extensively extended to incorporate a variety of coupled multiphysics formulations, including thermo‑mechanical \citep{Nairn2015, Tao2018, Zhao2022} and thermo‑hydro‑mechanical models \citep{Pinyol2018, Lei2021, yu2025}. Thermo‑mechanical MPM has been well employed to investigate the effects of temperature evolution on material behavior, such as thermal expansion and contraction \citep{Tao2018, Zhao2022}, thermal pressurization \citep{Pinyol2018, lei2024mpm}, thermally induced material softening \citep{yu2024, lei2025thm}, and heat generation due to plastic dissipation \citep{Nairn2015, yu2026fully} in large deformation, with applications spanning from permafrost mechanics \citep{yu2024c, yu2024d}, gas hydrate dissociation \citep{yu2026fully} to geothermal energy extraction. These scenarios involve tightly coupled thermal, hydraulic, and mechanical fields, imposing stringent requirements on the accuracy, robustness, and stability of numerical formulations. While substantial progress has been achieved in coupled multiphysics MPM, particularly through improvements in time‑integration schemes \citep{kularathna2021a, zheng2022} and spatial discretization strategies \citep{yamaguchi2021}, comparatively limited attention has been devoted to the accurate enforcement of boundary conditions \citep{coombs2023}. 

In the simulation of thermally coupled multiphysics problems, the accurate imposition of thermal boundary conditions is of fundamental importance, as the temperature field directly influences other physical fields through strong coupling mechanisms. Typical thermal boundary conditions include prescribed temperature (Dirichlet or essential boundary conditions) and prescribed heat flux (Neumann or natural boundary conditions). While temperature boundary conditions can usually be enforced in MPM with relative ease, the accurate application of heat flux boundary conditions is considerably more challenging. Nevertheless, heat flux boundaries are ubiquitous in both natural and engineering systems, as they govern energy exchange processes such as heat transfer between the atmosphere and the Earth’s surface, between flowing fluids and solid structures, and across material interfaces. Representative engineering applications include freeze–thaw cycles in permafrost and cryospheric soils \citep{hjort2022, luo2019}, as well as heat exchange in geothermal energy systems and geothermal piles \citep{tounsi2019, rad2016solar, bayer2014strategic}. In these problems, inaccuracies in the imposed thermal boundary conditions can propagate through the coupled system and significantly affect the hydraulic and mechanical responses.

\begin{figure}[!b]
    \centering
    \includegraphics[width=1\linewidth]{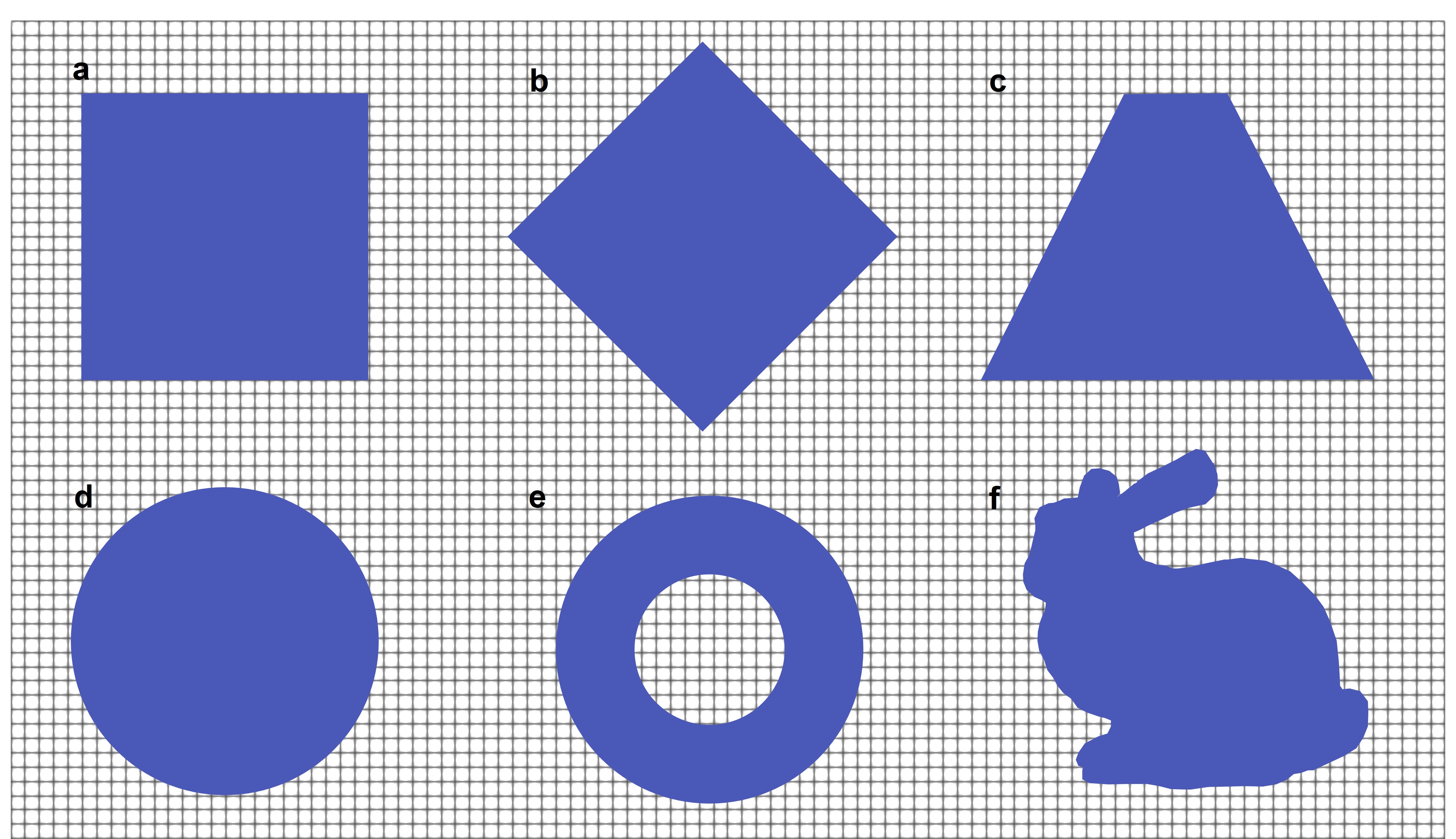}
    \caption{Illustration of material domains with varying geometries embedded in a regular background mesh. Except for the square domain in (a), all other geometries are partially or fully nonconforming with respect to the grid.}
    \label{fig: intro-1}
\end{figure}

The difficulty in enforcing heat flux boundary conditions in MPM primarily arises from the nonconforming relationship between material boundaries and the background grid \citep{liang2023b}. In MPM, a regular and fixed background mesh is typically employed, while materials are represented by a collection of Lagrangian material points whose geometry can be complex and evolve significantly due to large deformation and rotation. As a result, material boundaries are rarely aligned with grid lines, except in highly idealized cases. As illustrated in Fig. \ref{fig: intro-1}a, the square domain aligned with the grid can be treated accurately, whereas even a simple rotation renders the boundary nonconforming (see Fig. \ref{fig: intro-1}b). For more complex geometries, such as inclined or curved domains (e.g., Figs. \ref{fig: intro-1}c-e) or highly irregular shapes (e.g., the Stanford bunny in Fig. \ref{fig: intro-1}f \citep{curless1996volumetric}, boundary alignment with the regular background grid becomes practically impossible. Under such circumstances, directly applying heat flux conditions at grid nodes or material points leads to loss of accuracy and increased numerical error. Therefore, the development of robust and accurate techniques for imposing heat flux boundary conditions on nonconforming and evolving boundaries remains a critical and unresolved challenge in MPM simulations.

Although relatively few studies have explicitly focused on the imposition of heat flux boundary conditions in MPM, existing strategies for enforcing traction boundary conditions provide valuable references. One common approach is to apply boundary conditions directly to material points located near the boundary; however, because material points do not generally coincide with the exact geometric boundary and the effective thickness of the boundary layer used in integration is ill‑defined, this method often suffers from limited accuracy and poor convergence \citep{al2013formulation}. To improve boundary representation, some studies introduce virtual or auxiliary particles in the vicinity of the boundary, which can enhance accuracy but require additional particle generation and tracking procedures, thereby increasing algorithmic complexity and computational cost \citep{liang2024mortar}. Alternatively, moving or adaptive grid techniques have been proposed to better align the background grid with material boundaries, enabling more accurate boundary condition enforcement \citep{wang2017modelling}. Nevertheless, for complex or highly irregular geometries, achieving consistent boundary conformity remains difficult, particularly when regular background meshes are employed, which restricts the robustness and general applicability of these methods. 

All of the aforementioned approaches rely on additional tracking or reconstruction of boundary particles or grids. Although such strategies can improve accuracy, they inevitably increase algorithmic complexity and computational cost. Recently, a virtual stress boundary method was proposed to accurately impose traction boundary conditions in MPM without explicitly tracking material domain boundaries, offering a promising alternative perspective \citep{liang2023b, given2024virtual}. Inspired by this idea, the present study introduces a novel and computationally efficient Virtual Heat Flux Method (VHFM) for the accurate enforcement of Neumann boundary conditions on nonconforming boundaries. The central concept of VHFM is to construct a virtual heat flux field on the background grid that reproduces the prescribed boundary heat flux, thereby eliminating the need for explicit boundary tracking or boundary–grid alignment. On this basis, the governing equations are reformulated, and a unified expression for the virtual heat flux field is derived. The proposed method is not only suitable for heat‑conduction and thermo-mechanical problems but can also be naturally extended to more complex multiphysics applications.

The remainder of this paper is organized as follows. Section \ref{sec: governing equation} presents the governing heat‑transfer equations with heat flux boundary conditions and their discretizations within the MPM framework. Section \ref{sec: VHFM} introduces the Virtual Heat Flux Method (VHFM), including the underlying concept, the modified strong and weak formulations of the governing equations, and the corresponding solution algorithm. The accuracy and convergence properties of the proposed method are systematically verified through a series of benchmark problems in Section \ref{sec: Validations}. Finally, concluding remarks and perspectives for future work are provided in Section \ref{sec: Conclusions}.

\section{MPM for heat transfer problem}
\label{sec: governing equation}
In this section, we present the fundamental MPM framework for heat‑transfer analysis, including the governing equations, thermal boundary conditions, and their discretizations within the MPM formulation. We begin by introducing the mathematical notation adopted throughout this work. The symbol $\dot{\square}$ denotes the first‑order material time derivative. The operators $\square\cdot\square$ and $\square\colon\square$ represent the single and double tensor contraction, respectively, while $\square\otimes\square$ denotes the tensor product. Additionally, the subscripts $\square_p$ and $\square_I$ denote quantities associated with material points (also referred to as particles, indexed by $p$) and background grid nodes (indexed by $I$), respectively. Throughout this study, bold or blackboard bold notation is used to denote spatial variables in tensorial form.

\subsection{Heat transfer equation}
We consider transient heat transfer governed solely by thermal conduction within a material domain $\Omega \subset \mathbb{R}^d$ ($d = 2$ or $3$) (with boundary $ \partial \Omega $) over the time interval $\mathcal{T} = [0, t]$. In the absence of mechanical work and phase change, the local energy balance is expressed as \citep{tao2018b, zhao2020},
\begin{equation}
    \rho c \,\dot{T} + \nabla \cdot \boldsymbol{q} = Q 
    \quad \text{in } \Omega \times \mathcal{T}\, ,
\label{eq:heat_transfer}
\end{equation}
where $ T $ is the temperature, $ \rho $ is the material density, $ c $ is the specific heat capacity, $ \boldsymbol{q} $ is the conductive heat flux vector ($\rm W\cdot m^{-2}$), and $ Q $ denotes the volumetric heat source ($\rm W\cdot m^{-3}$).

Heat conduction is assumed to follow Fourier’s law,
\begin{equation}
    \boldsymbol{q} = - \boldsymbol\kappa \nabla T\, ,
\label{eq:fourier}
\end{equation}
where $ \boldsymbol\kappa $ is the thermal conductivity. In this study, heat transfer is assumed to be isotropic, and therefore $ \boldsymbol\kappa $ is reduced to a scalar constant $ \kappa $.

The boundary $ \partial \Omega $ is decomposed into two non‑overlapping parts, $ \partial \Omega_T $ and $ \partial \Omega_q $, on which temperature (Dirichlet) and heat flux (Neumann) boundary  conditions are prescribed, respectively. The thermal boundary conditions are defined as,
\begin{align}
    T &= \hat{T} 
    \quad &&\text{on } \partial \Omega_T \times \mathcal{T}\, , \label{eq:dirichlet_bc}\\
    - \boldsymbol{q} \cdot \boldsymbol{n} &= \hat{q} 
    \quad &&\text{on } \partial \Omega_q \times \mathcal{T}\, , \label{eq:neumann_bc}\\
    - \boldsymbol{q} \cdot \boldsymbol{n} &= \gamma (T - T_a) 
    \quad &&\text{on } \partial \Omega_q \times \mathcal{T}\, . \label{eq:convective_bc}
\end{align}
Here, $ \hat{T} $ denotes the prescribed temperature on $ \partial \Omega_T $, $ \hat{q} $ is the prescribed inward heat flux on $ \partial \Omega_q $, and $ \boldsymbol{n} $ is the outward unit normal vector to the boundary. Eq.~\eqref{eq:neumann_bc} represents a conductive heat flux boundary condition, while Eq.~\eqref{eq:convective_bc} describes convective heat exchange between the material surface and the surrounding environment, where $ \gamma $ is the convective heat‑transfer coefficient and $ T_a $ is the ambient temperature. The convective boundary condition describes surface heat exchange processes commonly encountered in natural and engineering systems, such as heat transfer between the ground surface and the atmosphere, or between solid structures and surrounding fluids. 

\subsection{Weak form and MPM discretization}

Let $\delta T$ denote an admissible virtual temperature field that satisfies the essential (Dirichlet) boundary conditions. 
Multiplying the heat transfer equation by $\delta T$, integrating over the domain $\Omega$, and applying integration by parts together with the divergence theorem, the weak form of the heat conduction equation is obtained as,
\begin{equation}
    \int_{\Omega} \rho c \,\dot{T}\, \delta T \, dV =
    \int_{\Omega} \boldsymbol{q} \cdot \nabla \delta T \, dV
    - \int_{\partial \Omega_q} \hat{q}\, \delta T \, dS
    + \int_{\Omega} Q\, \delta T \, dV \, ,
\label{eq:weak_form}
\end{equation}
where $\nabla \delta T$ denotes the gradient of the virtual temperature field.

In the MPM, the material domain is discretized by $n_p$ material points (particles), each carrying mass, volume, and state variables, and the volume integral is approximated by a summation over particles as follows,
\begin{equation}
    \int_{\Omega} (\cdot) dV = \sum_{p=1}^{n_p} (\cdot)V_p \, ,
\end{equation}
where $V_p$ is the particle volume. Therefore, the weak form Eq.~\eqref{eq:weak_form} can be discretized into a summation form,
\begin{equation}
    \sum_{p=1}^{n_p} V_p \rho_p c_p \dot{T}_p \, \delta T_p =
    \sum_{p=1}^{n_p} V_p \boldsymbol{q}_p \cdot \nabla \delta T_p
    + \sum_{p=1}^{n_p} V_p Q_p \, \delta T_p 
    - \int_{\partial \Omega_q} \hat{q}\, \delta T \, dS\, ,
\label{eq:weak_form_particle}
\end{equation}
where $\rho_p$, $c_p$, $T_p$, $\boldsymbol{q}_p$, and $Q_p$ denote the particle density, heat capacity, temperature, conductive heat flux, and volumetric heat source, respectively. If no prescribed heat flux is applied on the boundary (i.e., $\hat{q} = 0$), the boundary integral vanishes.

The particle temperature $T_p$ and its virtual counterpart $\delta T_p$ are interpolated from the background grid nodal values using MPM shape functions $S_{Ip}$ as,
\begin{equation}
T_p = \sum_{I=1}^{n_n} S_{Ip} T_I\, ,
\qquad
\delta T_p = \sum_{I=1}^{n_n} S_{Ip} \delta T_I\, ,
\label{eq:temperature_interpolation}
\end{equation}
where $T_I$ and $\delta T_I$ are the nodal temperature and virtual temperature at node $I$, and $n_n$ is the number of grid nodes.

Substituting Eq.~\eqref{eq:temperature_interpolation} into Eq.~\eqref{eq:weak_form_particle} and exploiting the arbitrariness of $\delta T_I$ leads to the semi‑discrete nodal temperature equation,
\begin{equation}
\mathcal{C}_I \, \dot{T}_I
=
\mathcal{E}_I^{\mathrm{int}}
+
\mathcal{E}_I^{\mathrm{ext}}\, ,
\label{eq:nodal_temperature}
\end{equation}
where the nodal heat capacity, internal heat contribution, and external heat contribution are defined as,
\begin{align}
\mathcal{C}_I &= \sum_{p=1}^{n_p} V_p \rho_p c_p S_{Ip}\, , \label{eq:capacity_matrix}\\
\mathcal{E}_I^{\mathrm{int}} &= \sum_{p=1}^{n_p} V_p \boldsymbol{q}_p \cdot \nabla S_{Ip}\, , \label{eq:internal_heat}\\
\mathcal{E}_I^{\mathrm{ext}} &= \sum_{p=1}^{n_p} V_p Q_p S_{Ip}
- \int_{\partial \Omega_q} \hat{q} \, S_I \, dS\,  . \label{eq:external_heat}
\end{align}

In Eq.~\eqref{eq:internal_heat}, the conductive heat flux at each particle is evaluated from the interpolated temperature gradient according to Fourier’s law,
\begin{equation}
\boldsymbol{q}_p
=
- \kappa \nabla T_p
=
- \kappa \sum_{I=1}^{n_n} T_I \nabla S_{Ip}\, .
\label{eq:particle_heat_flux}
\end{equation}

The explicit forward Euler integration scheme adopted for the temporal discretization of the transient heat equation is conditionally stable. To prevent numerical oscillations and ensure a converged solution, the time step $\Delta t$ must be constrained by the critical time step $\Delta t_{\text{cr}}$, derived from the Courant–Friedrichs–Lewy (CFL) condition adapted for diffusion problems,
\begin{equation}
    \Delta t \le \Delta t_{\text{cr}} = \frac{h_{\min}^2}{\alpha}\,,
\end{equation}
where $h_{\min}$ denotes the minimum mesh size and $\alpha = k/(\rho c)$ is the thermal diffusivity. 


\subsection{Conventional heat flux boundary imposition}
As indicated by the weak form in Eq.~\eqref{eq:weak_form}, the enforcement of heat flux boundary conditions requires the evaluation of a surface integral over the Neumann boundary $\partial \Omega_q$. In the finite element method (FEM), this operation is straightforward because mesh nodes and element faces are explicitly aligned with the material boundary, allowing direct numerical integration using boundary elements. In the MPM, however, as aforementioned, the background grid usually does not conform to the evolving material boundary. A commonly adopted strategy to impose heat flux boundary conditions in MPM is the use of \emph{boundary particles}. 
These particles constitute a subset of material points that are identified as being located in the vicinity of the boundary $\partial \Omega_q$. The prescribed heat flux is applied to these boundary particles, and the resulting thermal contribution is converted into nodal quantities.

Under this approach, the boundary integral of the prescribed heat flux is approximated as,
\begin{equation}
    \int_{\partial \Omega_q} \hat{q}S_{I} \, dS
    \approx
    \sum_{p=1}^{n_p} A_p \hat{q}_pS_{Ip}
    =
    \sum_{p=1}^{n_p} V_p h_p^{-1} \hat{q}_pS_{Ip}\,,
\label{eq:boundary_particle_flux}
\end{equation}
where $A_p$ denotes the effective surface area associated with a boundary particle, $V_p$ is the particle volume, and $h_p$ represents an assumed characteristic thickness of the boundary particle normal to the surface. Despite its conceptual simplicity, this conventional boundary‑particle‑based approach suffers from several inherent limitations. First, the definition of the particle surface area $A_p$ and the associated thickness $h_p$ is ambiguous, particularly for irregular geometries or curved boundaries.  Second, the identification of boundary particles is heuristic and mesh‑dependent, which may lead to inconsistent boundary representation as the material undergoes large deformation or particle redistribution. 


Alternatively, heat flux boundary conditions may be imposed directly at the grid‑nodal level without introducing boundary particles. In this nodal‑based approach, the surface integral of the prescribed heat flux is approximated as,
\begin{equation}
    \int_{\partial \Omega_q} \hat{q} S_{I}\, dS
    \approx
    \sum_{I=1}^{n_I} A_I \hat{q}_IS_{I}\,,
\label{eq:nodal_boundary_flux}
\end{equation}
where $A_I$ denotes the effective boundary surface area associated with node $I$, and $\hat{q}_I$ is the prescribed heat flux applied at that node. This approach requires the explicit identification of boundary nodes and the assignment of appropriate nodal surface areas. Similarly, as the background grid in MPM does not conform to the material boundary, the determination of $A_I$ is non‑unique and generally mesh‑dependent. 

These drawbacks of conventional heat flux boundary imposition methods motivate the development of alternative boundary‑treatment strategies that avoid explicit surface reconstruction and provide a more robust and accurate enforcement of heat flux boundary conditions within the MPM framework.

\section{Virtual heat flux method (VHFM)}
\label{sec: VHFM}
The Virtual Heat Flux Method (VHFM) is proposed to impose heat flux boundary conditions in the MPM without explicit boundary reconstruction, boundary particles, or nodal surface integration. The key idea is to replace the surface heat flux contribution by an equivalent volumetric contribution defined on an auxiliary \emph{virtual domain}.

\subsection{Virtual domain and virtual heat flux}
\begin{figure}[!htb]
\centering
\includegraphics[width=0.85\linewidth]{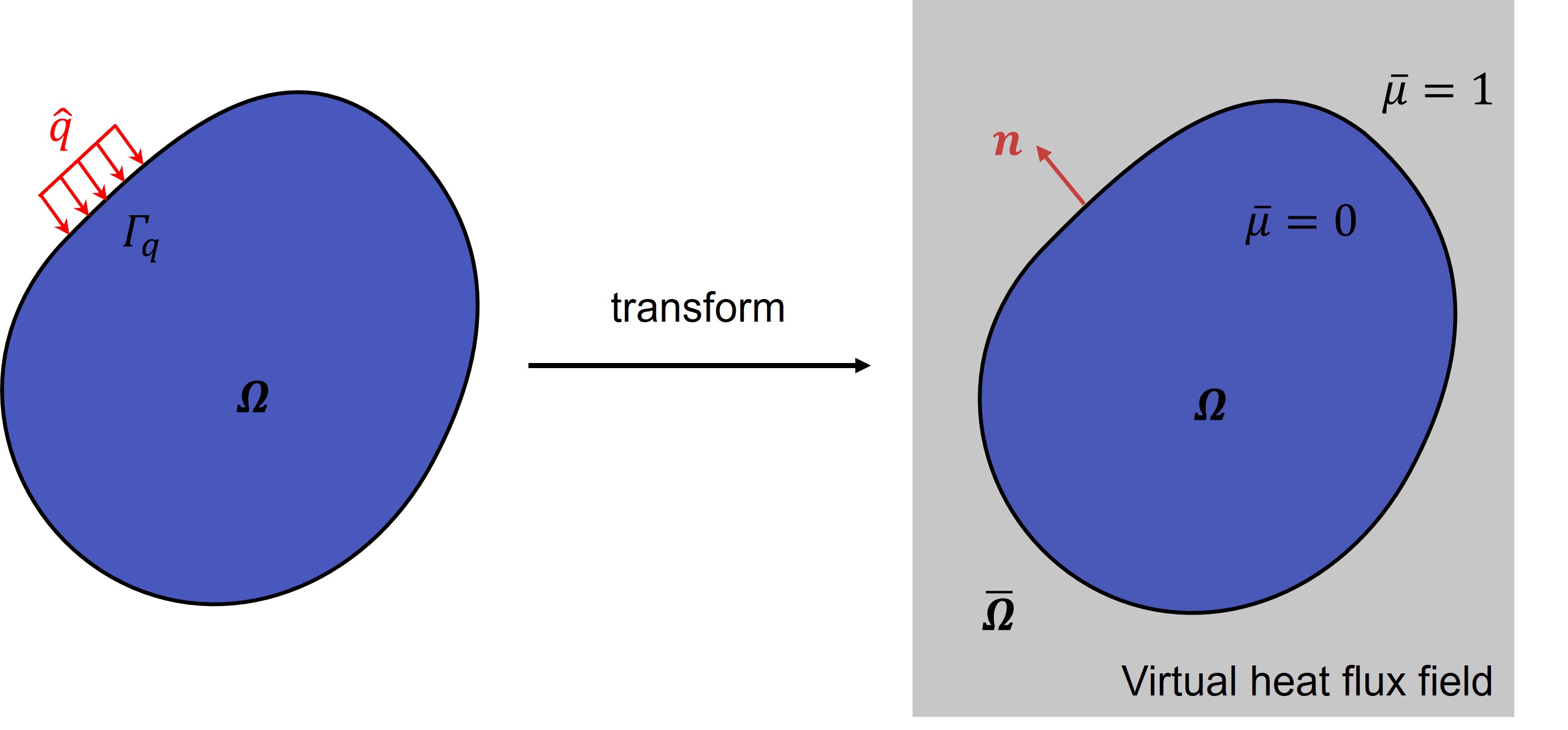}
\caption{Schematic illustration of the virtual domain $\bar{\Omega}$ surrounding the physical domain $\Omega$ and the associated virtual heat flux field.}
\label{fig:vhfm_domain}
\end{figure}

We introduce an auxiliary \emph{virtual domain} $\bar{\Omega}$ such that its intersection with the physical domain $\Omega$ coincides with the material boundary,
\begin{equation}
\bar{\Omega} \cap \Omega = \partial \Omega \, .
\label{eq:virtual_domain}
\end{equation}
The virtual domain may be chosen as an arbitrary layer that surrounds the material boundary, as illustrated in Fig.~\ref{fig:vhfm_domain}. It does not represent a physical region of the body, but rather serves as a mathematical construct that enables the reformulation of surface boundary contributions into equivalent volumetric terms.

Within $\bar{\Omega}$, a virtual heat flux field $\bar{\boldsymbol{q}}$ is defined to satisfy the prescribed heat flux boundary condition,
\begin{equation}
\bar{\boldsymbol{q}} \cdot \boldsymbol{n} =
\begin{cases}
\hat{q}, & \text{on } \partial \Omega_q\,, \\
0, & \text{on } \partial \Omega \setminus \partial \Omega_q \,,
\end{cases}
\label{eq:virtual_flux_bc}
\end{equation}
This construction ensures that the prescribed heat flux is enforced exactly in the normal direction, while naturally vanishing on boundaries where no flux is applied.

In addition, the virtual heat flux field is required to satisfy a homogeneous heat‑transfer equation within the virtual domain,
\begin{equation}
\rho c \, \dot{\bar{T}} + \nabla \cdot \bar{\boldsymbol{q}} = 0
\quad \text{in } \bar{\Omega} \,,
\label{eq:virtual_governing}
\end{equation}
where $\bar{T}$ is an auxiliary temperature field defined solely for theoretical consistency. 
Importantly, $\bar{T}$ is not a physical temperature and does not need to be solved explicitly. 
Its sole purpose is to ensure that the virtual heat flux field admits a divergence form compatible with the governing equation, thereby allowing the boundary heat flux contribution to be incorporated into the weak form as a volumetric term.

Conceptually, the virtual domain and its associated fields act as a mathematical device that ``absorbs'' the boundary heat flux and redistributes it into the interior weak form. As a result, the original surface integral associated with the Neumann boundary condition is replaced by an equivalent volume integral involving $\bar{\boldsymbol{q}}$, which can be evaluated robustly using standard MPM particle integration. This transformation is the central idea underlying the VHFM.

\subsection{Modified governing equation and MPM discretization}
To distinguish quantities defined in the physical and virtual domains, we introduce an indicator function $\bar{\mu}$,
\begin{equation}
\bar{\mu} =
\begin{cases}
1, & \text{in } \bar{\Omega}\,, \\
0, & \text{in } \Omega \, .
\end{cases}
\end{equation}
Using this indicator, the governing equations in the physical and virtual domains may be expressed in a unified form over the extended domain $\Omega \cup \bar{\Omega}$ as
\begin{equation}
(1-\bar{\mu}) \left( \rho c \dot{T} + \nabla \cdot \boldsymbol{q} - Q \right)
+ \bar{\mu} \left( \rho c \dot{\bar{T}} + \nabla \cdot \bar{\boldsymbol{q}} \right)
= 0 \, .
\end{equation}
This combined expression serves only as an intermediate mathematical device that allows the boundary contribution to be embedded into a volumetric formulation.

Invoking the homogeneous governing equation in the virtual domain, Eq.~\eqref{eq:virtual_governing}, and restricting the resulting expression to the physical domain $\Omega$, one obtains
\begin{equation}
\rho c \left( \dot{T} - \dot{\bar{T}} \right)
+ \nabla \cdot \left( \boldsymbol{q} - \bar{\boldsymbol{q}} \right)
- Q
= 0
\quad \text{in } \Omega \, .
\end{equation}
Since the auxiliary temperature field $\bar{T}$ is defined only to ensure variational consistency and is not solved explicitly, the difference $(\dot{T}-\dot{\bar{T}})$ may be interpreted as an effective temperature rate that accounts for the presence of the prescribed heat‑flux boundary condition.

For explicit time integration, this effective rate is approximated by the forward temperature rate at the next time step, yielding the modified strong form
\begin{equation}
\rho c \dot{T}^{k+1}
+ \nabla \cdot \left( \boldsymbol{q}^k - \bar{\boldsymbol{q}}^{k+1} \right)
- Q
= 0
\quad \text{in } \Omega \, .
\label{eq:vhfm_strong}
\end{equation}
This expression differs from the classical heat equation only through the additional divergence term associated with the virtual heat flux, which encapsulates the Neumann boundary condition.

The weak form is obtained by multiplying Eq.~\eqref{eq:vhfm_strong} by an admissible virtual temperature $\delta T$ and integrating over the physical domain,
\begin{equation}
\int_{\Omega} \rho c \, \delta T \, \dot{T}^{k+1} \, dV
- \int_{\Omega} \left( \boldsymbol{q}^k - \bar{\boldsymbol{q}}^{k+1} \right) \cdot \nabla \delta T \, dV
- \int_{\Omega} \delta T \, Q \, dV
= 0 \, .
\label{eq:vhfm_weak}
\end{equation}
Compared to the conventional weak form, the boundary integral associated with the prescribed heat flux does not appear explicitly. Its effect is instead incorporated through the volumetric term involving $\bar{\boldsymbol{q}}$, which arises naturally from the divergence theorem.

Applying standard MPM discretization, the temperature field and test function are approximated using nodal shape functions. Substitution into Eq.~\eqref{eq:vhfm_weak} leads to the semi‑discrete nodal temperature equation
\begin{equation}
\mathcal{C}_{IJ} \dot{T}_J
=
\widetilde{\mathcal{E}}_I^{\mathrm{int}}
+
\widetilde{\mathcal{E}}_I^{\mathrm{ext}} \, ,
\end{equation}
where $\mathcal{C}_{IJ}$ is the consistent nodal heat‑capacity matrix. The internal and external thermal force vectors are given by
\begin{align}
\widetilde{\mathcal{E}}_I^{\mathrm{int}}
&=
\sum_{p=1}^{n_p} V_p \, \boldsymbol{q}_p \cdot \nabla S_{Ip}
-
\sum_{p=1}^{n_p} V_p \, \bar{\boldsymbol{q}}_p \cdot \nabla S_{Ip} \, , \\
\widetilde{\mathcal{E}}_I^{\mathrm{ext}}
&=
\sum_{p=1}^{n_p} V_p \, Q_p \, S_{Ip} \, .
\end{align}

Notably, the conventional surface integral over the Neumann boundary $\partial \Omega_q$ is completely eliminated. 
The prescribed heat flux is enforced instead through the volumetric contribution associated with the virtual heat‑flux field. This reformulation yields a boundary treatment that is fully compatible with the particle integration framework of MPM, avoids explicit surface discretization, and remains robust under large deformations and evolving boundaries.

\subsection{Choice of virtual heat flux field}
Any virtual heat flux field that satisfies Eq.~\eqref{eq:virtual_flux_bc} may, in principle, be adopted. The virtual field is not unique, and different choices may be constructed provided that the prescribed normal flux on the boundary is recovered. In this work, a particularly simple and robust choice is employed,
\begin{equation}
\bar{\boldsymbol{q}} = \hat{q} \, \boldsymbol{n} \, .
\label{eq:virtual_flux_definition}
\end{equation}
where $\boldsymbol{n}$ is the outward unit normal vector of the physical domain $\Omega$. 

This choice possesses several desirable properties. First, it satisfies the prescribed heat flux boundary condition exactly,
\begin{equation}
\bar{\boldsymbol{q}} \cdot \boldsymbol{n}
= \hat{q} \, \boldsymbol{n} \cdot \boldsymbol{n}
= \hat{q} \, .
\end{equation}
Second, the virtual heat flux is purely normal to the boundary, ensuring that no spurious tangential flux components are introduced. Third, the definition of $\bar{\boldsymbol{q}}$ requires only the outward unit normal vector, making it straightforward to implement.

\begin{remark}
   It is emphasized that the virtual heat flux field is not intended to represent the physical heat flux within the body. Instead, it acts as an auxiliary field whose divergence generates an equivalent volumetric contribution in the weak form, replacing the original surface heat flux integral. This non‑uniqueness provides flexibility, but the present choice is preferred due to its minimal complexity and consistent performance. 
\end{remark}

\begin{remark}
The VHFM formulation is not limited to thermal problems but can be applied directly to general Neumann‑type boundary conditions in MPM. Depending on the nature of the prescribed boundary quantity, the corresponding virtual field may be constructed as follows.

\paragraph{Scalar Neumann boundary conditions}
For a prescribed scalar flux $\hat{j}$ (e.g., heat flux or fluid flux),
\begin{equation}
\boldsymbol{J} \cdot \boldsymbol{n} = \hat{j} \quad \text{on } \partial \Omega_N ,
\end{equation}
a virtual vector field may be defined as
\begin{equation}
\bar{\boldsymbol{J}} = \hat{j} \, \boldsymbol{n} \, ,
\end{equation}
which guarantees
$\bar{\boldsymbol{J}} \cdot \boldsymbol{n} = \hat{j}$.

\paragraph{Vector Neumann boundary conditions}
For a prescribed vector boundary traction $\hat{\boldsymbol{t}}$,
\begin{equation}
\boldsymbol{\sigma} \cdot \boldsymbol{n} = \hat{\boldsymbol{t}} \quad \text{on } \partial \Omega_t \,,
\end{equation}
a corresponding second‑order virtual stress tensor may be introduced as
\begin{equation}
\bar{\boldsymbol{\sigma}} = \hat{\boldsymbol{t}} \otimes \boldsymbol{n} \, ,
\end{equation}
which satisfies
\begin{equation}
\bar{\boldsymbol{\sigma}} \cdot \boldsymbol{n}
= (\hat{\boldsymbol{t}} \otimes \boldsymbol{n}) \cdot \boldsymbol{n}
= \hat{\boldsymbol{t}} \, .
\end{equation}

\paragraph{Tensor Neumann boundary conditions}
More generally, for higher‑order or tensor‑valued Neumann boundary conditions, a virtual field of appropriate order may be constructed by taking the tensor product of the prescribed boundary quantity with the outward normal vector. 
This construction ensures that the contraction of the virtual field with the normal vector recovers the prescribed boundary condition exactly.

These constructions highlight that VHFM provides a unified and systematic framework for imposing Neumann boundary conditions of different tensorial orders, while maintaining a purely volumetric weak formulation.
\end{remark}

\subsection{Surface-normal estimation via a scalar field}
\label{subsec:normal_estimation}
The surface normal can be evaluated using the mass‑gradient method, which is simple and widely adopted in MPM. The nodal mass field exhibits a sharp variation near the material boundary, and its gradient provides a natural approximation of the outward normal direction. This approach avoids additional boundary reconstruction and fits naturally within the standard MPM formulation. For isotropic materials, the outward unit normal at particle $p$ is calculated as,
\begin{equation}
\boldsymbol{n}_p = \frac{\nabla m_p}{\| \nabla m_p \|} \, ,
\label{eq:normal_estimation}
\end{equation}
with
\begin{equation}
\nabla m_p = \sum_{I=1}^{n_I} m_I \nabla S_{Ip} \, ,
\end{equation}
where $m_I$ is the nodal mass and $S_{Ip}$ denotes the MPM shape function. However, the mass gradient method may lose accuracy when the material density is inhomogeneous. 

Alternatively, the outward unit normal vector required can be evaluated by introducing a \textit{unified scalar field} $\phi$ that takes a constant value (e.g., $\phi = 1$) for all material points inside the physical domain $\Omega$ and is zero outside. The gradient of this field naturally points in the direction of the steepest ascent of the indicator function, which coincides with the outward normal direction at the boundary.

The nodal values of the scalar field are obtained through the standard MPM mapping,
\begin{equation}
\phi_I = \sum_{p} S_{Ip} \, \phi_p \, V_p,
\end{equation}
where $\phi_p = 1$ for all material points. The gradient at a material point is then computed as,
\begin{equation}
\nabla \phi_p = \sum_{I} \phi_I \, \nabla S_{Ip}.
\end{equation}

Finally, the unit outward normal is given by,
\begin{equation}
\boldsymbol{n}_p = \frac{\nabla \phi_p}{\|\nabla \phi_p\|}.
\end{equation}
This \textit{constant-scalar-gradient method} decouples the normal estimation from the material's density distribution. It is therefore robust in problems involving large density contrasts, non-uniform mass distributions, or purely geometric models where mass is not defined. 

\begin{figure}[!htb]
\centering
\includegraphics[width=0.55\linewidth]{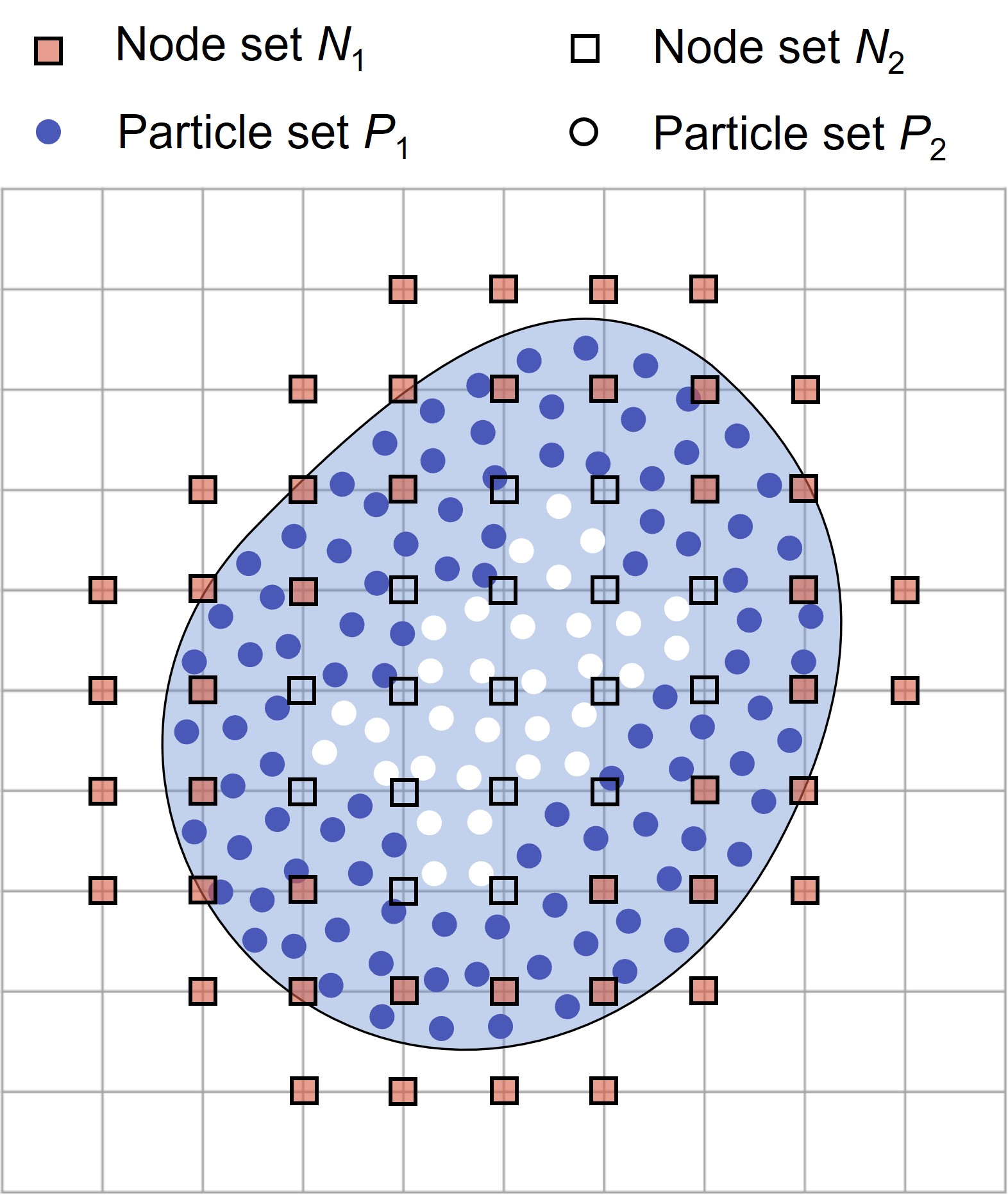}
\caption{Conceptual illustration of surface nodes and surface particles. $P_1$: surface particle set; $P_2$: non-surface particle set; $N_1$: surface node set; $P_2$: non-surface node set.}
\label{fig: vhfm_concept}
\end{figure}

\subsection{Solution algrithms}
The numerical implementation of the VHFM is given below and summarized in Algorithm~\ref{alg:vfm-complete}.

\begin{algorithm}[!htb]
\caption{Virtual Heat Flux Method for Thermal Boundary Condition Impositions}
\label{alg:vfm-complete}
\begin{algorithmic}[1]
\Require 
\State \textbf{Step 1: Identify free surface nodes}
\State $\mathcal{N}_1 \gets \emptyset$
\State $N_c \gets$ number of active cells; $V_{p\in cJ} \gets$ total volume of the particles in cell $J$; $V_{J} \gets$ volume of the cell $J$; 
\For{$J = 1$ to $N_c$}
    \If{$0 < V_{p\in cJ} / V_{cJ} < \eta$} \Comment{Volume fraction criterion}
        \State $\mathcal{N}_1 \gets \mathcal{N}_{\text{1}} \cup \{\text{Nodes} \in J\}$
    \EndIf
\EndFor

\medskip
\State \textbf{Step 2: Compute surface normals via scalar gradient}
\State $n_p \gets$ number of particles
\For{each node $p = 1$ to $n_p$}
    \State $n_{n} \gets$ number of nodes connected to particle $p$
    \For{each node $I = 1$ to $n_{n}$}
        \State $\phi_I \gets \phi_p S_{Ip}$
    \EndFor
\EndFor

\For{each particle $p = 1$ to $n_p$}
    \State Compute scalar gradient: $\nabla \phi_p = \sum_{I} \phi_I \nabla S_{Ip}$
    \State Compute unit normal: $\boldsymbol{n}_p = \dfrac{\nabla \phi_p}{\|\nabla \phi_p\|}$
\EndFor

\medskip
\State \textbf{Step 3: Impose boundary flux via VHFM}
\For{each particle $p = 1$ to $n_p$}
    \For{each node $I = 1$ to $n_{n}$}
        \If{$I \in \mathcal{N}_1$}
            \State $\mathcal{E}_I^{\text{int}} \gets \mathcal{E}_I^{\text{int}} + V_p \, \hat{q}_p \, \boldsymbol{n}_p \cdot \nabla S_{Ip}$
        \EndIf
    \EndFor
\EndFor

\State \Return Updated $\mathcal{E}_I^{\text{int}}$
\end{algorithmic}
\end{algorithm}

\begin{enumerate}
    \item \textit{Boundary node detection}. Boundary nodes are identified via the volume fraction method. If the total particle volume in the cell (denoted as $\mathcal{V}_{p\in c}$) relative to the cell volume (denoted as $\mathcal{V}_c$) falls below a prescribed threshold $\eta \in (0,1)$, all nodes for the cell are classified as free surface nodes (node set $\mathcal{N}_1$ in Fig.~\ref{fig: vhfm_concept}). By default, $\eta$ is taken as 0.5. The corresponding particles associated with these boundary nodes are subsequently assigned as boundary particles (particle set $\mathcal{P}_2$ in Fig.~\ref{fig: vhfm_concept}). Other free surface detection methods can also be adopted \citep{liang2023b, given2024virtual}.

    \item \textit{Computation of boundary particle outward unit normal vector}. The unit outward normal vector at each boundary particle is computed using the mass gradient method. The particle mass gradient is evaluated through the summation of nodal mass contributions weighted by the shape function gradients, followed by normalization to obtain the unit normal (Eq.~\eqref{eq:normal_estimation}). 
    
    \item \textit{Imposition of heat‑flux boundary conditions}. After identifying the free‑surface nodes and computing the corresponding particle outward normals, the prescribed heat‑flux boundary condition is enforced in weak form using the virtual flux formulation. The virtual heat flux defined at boundary particles is mapped to the nodal internal heat exclusively at free‑surface nodes.
\end{enumerate}

\section{Numerical examples}
\label{sec: Validations}
In this section, the performance of the proposed VHFM for heat flux boundary imposition in the MPM is assessed. Five numerical examples have been conducted with different material geometries and heat flux boundaries, each with a specific purpose:
\begin{enumerate}
    \item Firstly, a 1D semi-infinite rod test with constant and convective flux boundaries is conducted, aiming to verify the fundamental accuracy and spatial convergence of the VHFM by direct comparison with analytical results.

    \item Secondly, the heating of a 2D circular ring (with both concave and convex boundaries) is simulated, aiming to demonstrate the method's capability for handling complex static geometries without boundary conformity, verified against reference finite difference solutions.

    \item Thirdly, the convective cooling of a 3D sphere is simulated to demonstrate the method's effectiveness in modeling 3D scenarios. The effect of particles per cell on the accuracy is also investigated. 

    \item Moreover, the heating of a rotating 2D square block is simulated, involving both fixed-angle and continuously rotating configurations, aiming to validate the method's accuracy and robustness for dynamically evolving nonconforming boundaries under rigid-body motion. 

    \item Finally, the method is applied to a practical problem, the cooling of a rotating fan with complex geometry, aiming to showcase the method's robustness and applicability in simulating heat transfer processes with intricate, moving boundaries.  
\end{enumerate}

\subsection{1D example: Transient Heat transfer in a semi-infinite Rod}
\label{sec: Heating of 1D semi-infinite Rod}
\begin{figure}[!htb]
    \centering
    \includegraphics[width=1.0\linewidth]{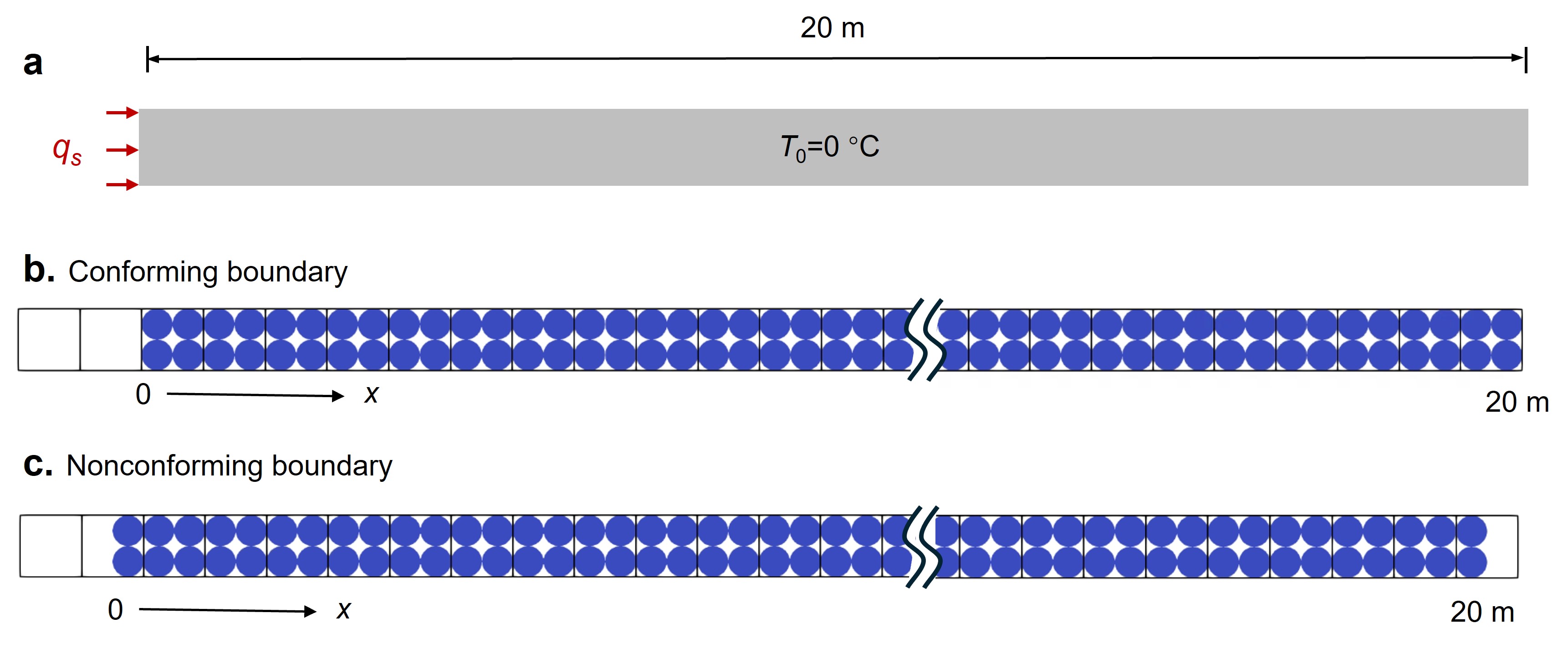}
    \caption{Transient heat transfer in a 1D semi-infinite Rod: (a) model geometry, initial and boundary conditions, (c) conforming boundary configuration, and (b) nonconforming boundary configuration.}
    \label{fig:1D-rod}
\end{figure}

In the first example, we simulate the transient heat transfer problem of a semi-infinite rod under a heat flux boundary. This problem can be simplified to a one-dimensional (1D) scenario, as illustrated in Fig. \ref{fig:1D-rod}a, where the left boundary is subjected to a heat flux $q_s$. To approximate a semi-infinite condition, the rod length, $L$, is set to a sufficiently large size of 20 m, and we focus only on the results within the first 5 m region. The simulation is conducted using a unity material assumption, where all material properties, including the density $\rho$ ($\rm kg/m^3$), the specific heat capacity $c$ ($\rm J/(kg\cdot{^\circ}{C})$), and the thermal conductivity $\kappa$ ($\rm W/(m\cdot{^\circ}{C})$), are set to 1. The initial temperature, $T_0$, is set to 0°C. We examined both a constant heat flux boundary and a convective heat flux boundary condition. For the former, the applied heat flux $q_s$ is set to 1 $\rm W/m^2$. For the latter, the ambient temperature, $T_a$, is set to 1°C, and the convective heat transfer coefficient, $\gamma$, is also set to 1 $\rm W/(m^2\cdot{^\circ}C)$. 

The analytical solution for the temperature at location $x$ and time $t$ is expressed as follows:
\begin{itemize}
    \item For constant heat flux boundary: 
\end{itemize}
\begin{gather}
    T(x, t) = T_0 + 2 \frac{q_s}{\kappa} \sqrt{\frac{\alpha t}{\pi}}
    \left[
    \exp\left( \frac{-x^2}{4 \alpha t} \right)
    - \frac{1}{2}x \sqrt{\frac{\pi}{\alpha t}} , \text{erfc}\left( \frac{x}{2\sqrt{\alpha t}} \right)
    \right]\,,
\label{eq: const heat flux boundary}
\end{gather}

\begin{itemize}
    \item For convective heat flux boundary: 
\end{itemize}
\begin{equation}
    T(x, t) = T_0 + (T_a - T_0)
    \Bigg[ 
    \text{erfc} \left( \frac{x}{2 \sqrt{\alpha t}} \right)  \\
     - \exp\left( \frac{\gamma x}{\kappa} + \frac{\gamma^2 \alpha t}{\kappa^2} \right)
    \cdot \text{erfc} \left( \frac{x}{2 \sqrt{\alpha t}} + \frac{\gamma \sqrt{\alpha t}}{\kappa} \right)
    \Bigg]\,,
\label{eq: convective heat flux boundary}
\end{equation}
where $\alpha = \kappa/(\rho c)$ is the thermal diffusivity and $\text{erfc}$ is the complementary error function. The detailed derivation of the above analytical solution is presented in \ref{app:1d_analytic}

A background grid with a cell size of 0.1 m is employed for the simulation. The material point size is set to half of the grid size, i.e., 0.05 m. Two different arrangements of material points are considered, as illustrated in Fig.~\ref{fig:1D-rod}. In Fig.~\ref{fig:1D-rod}b, the material points are arranged in a \textit{conforming boundary} configuration, where each grid cell contains four evenly distributed material points, and the material points at the left boundary align with the grid boundary. In contrast, Fig.~\ref{fig:1D-rod}c shows a \textit{nonconforming boundary} configuration, where all material points are shifted leftward by half a grid size. In this case, the leftmost grid cell contains only two material points. The VHFM is employed to test both the conforming and nonconforming boundary configurations. For comparison, in the conforming boundary configuration, two additional simulations are conducted by applying the boundary conditions either directly at the nodes or at the material points. A uniform time step of $\Delta t = 1\times10^{-3}$ s is used for the simulations.

\begin{remark}
    Note that the material points in our simulations are arranged in the commonly used equidistant configuration, located at positions 0.25 $h$ and 0.75 $h$ within each grid cell, rather than being precisely positioned at the Gaussian integration points. While this arrangement may introduce additional integration errors, these errors are consistent across all cases and therefore do not affect the validity of our conclusions.
\end{remark}

\begin{figure}[!htb]
    \centering
    \includegraphics[width=1\linewidth]{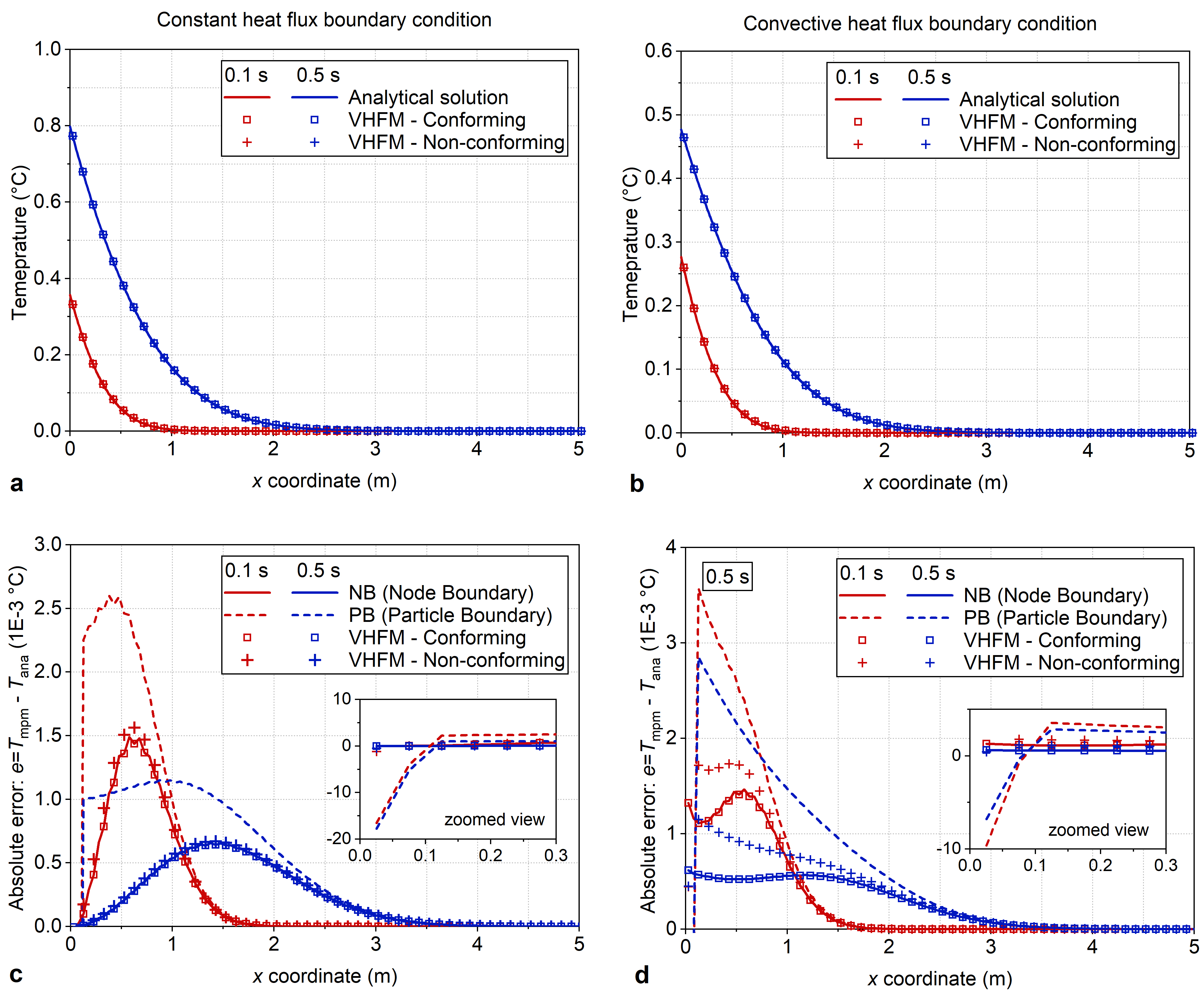}
    \caption{Transient heat transfer in a 1D semi-infinite Rod with both constant heat flux boundary (left column) and convective heat flux boundary (right column): (a-b) comparison of the temperature distributions at time instants of 0.1 s and 0.5 s obtained using MPM with VHFM and the analytical solution; (c-d) comparison of the absolute error for different boundary condition imposition methods. The zoomed-in views in (c) and (d) highlight the error near the origin, i.e., at the heat flux boundary, indicating that applying heat flux directly at the particle boundary results in significant errors near the boundary.}
    \label{fig: ex1-2}
\end{figure}
Figs.~\ref{fig: ex1-2}a and b compare the temperature distributions at time instants $t = 0.1\ \mathrm{s}$ and $t = 0.5\ \mathrm{s}$, obtained using the MPM with VHFM and the analytical solution. It can be observed that for both constant and convective heat flux boundaries, the temperature fields simulated using VHFM agree very well with the theoretical solution, regardless of whether the boundary is conforming or nonconforming.

To quantitatively analyze the simulation error, the absolute error, $e_p = T_{\text{mpm},\,p} - T_{\text{ana},\,p}$, is plotted in Figs.~\ref{fig: ex1-2}c and d, where $T_{\text{mpm},\,p}$ and $T_{\text{ana},\,p}$ represent the simulated and theoretical temperatures at material point $p$, respectively. Additionally, the root mean square error (RMSE), $\varepsilon_{\text{err}}$, between the temperature field obtained from the MPM simulation and the analytical solution is calculated,
\begin{equation}
    \varepsilon_{\text{err}} = \sqrt{\frac{1}{N_p} \sum_{p=1}^{N_p} (T_{\text{mpm},\,p} - T_{\text{ana},\,p})^2}\,,
    \label{eq:placeholder}
\end{equation}
where $N_p$ is the total number of material points.

\begin{table}[!htb]
    \caption{Transient heat transfer in a 1D semi-infinite Rod: comparison of RMSE $\varepsilon_{err}$ using different boundary condition imposition methods.}
    \centering
    {
    \footnotesize
    \begin{tabular*}{\textwidth}{@{\extracolsep{\fill}}ccccc@{}}
        \toprule
        \multicolumn{5}{c}{Constant heat flux boundary condition} \\
        \midrule
        Time & Node boundary (NB) & Particle boundary (PB) & VHFM-Conforming & VHFM-nonconforming  \\
        \midrule
        0.1 s & $2.420\times 10^{-4}$ & $9.714\times 10^{-4}$ & $2.420\times 10^{-4}$ & $2.599\times 10^{-4}$ \\
        0.5 s & $1.608\times 10^{-4}$ & $9.803\times 10^{-4}$ & $1.608\times 10^{-4}$ & $1.661\times 10^{-4}$ \\
        1.0 s & $1.352\times 10^{-4}$ & $9.826\times 10^{-4}$ & $1.352\times 10^{-4}$ & $1.383\times 10^{-4}$ \\
        \midrule
        \midrule
        \multicolumn{5}{c}{Convective heat flux boundary condition} \\
        \midrule
        Time & Node boundary (NB) & Particle boundary (PB) & VHFM-Conforming & VHFM-nonconforming  \\
        \midrule
        0.1 s & $2.810\times 10^{-4}$ & $7.173\times 10^{-4}$ & $2.810\times 10^{-4}$ & $2.065\times 10^{-4}$ \\
        0.5 s & $1.748\times 10^{-4}$ & $6.159\times 10^{-4}$ & $1.748\times 10^{-4}$ & $2.574\times 10^{-4}$ \\
        2.5 s & $8.463\times 10^{-5}$ & $5.756\times 10^{-4}$ & $8.463\times 10^{-5}$ & $3.392\times 10^{-4}$ \\
        \bottomrule
    \end{tabular*}
    }
\label{table: ex1-1}
\end{table}

Figs.~\ref{fig: ex1-2}c and d further compare the absolute errors obtained using the VHFM and the conventional particle and node boundary approaches (both under the conforming condition). Table~\ref{table: ex1-1} summarizes the calculated $\varepsilon_{\text{err}}$ at different time instants for various boundary imposition methods. The results show that for conforming boundaries, the accuracy of the VHFM simulation is identical to that of directly applying the boundary conditions at the nodes, both of which are significantly more accurate than applying the boundary conditions at the particles. For nonconforming boundaries, the accuracy of VHFM is slightly lower but remains close to that of the conforming condition with node-based boundary imposition, and it is still far more accurate than the particle-based approach. These findings demonstrate that the proposed VHFM achieves excellent accuracy when imposing heat flux boundaries.

\begin{figure}[!htb]
    \centering
    \includegraphics[width=1\linewidth]{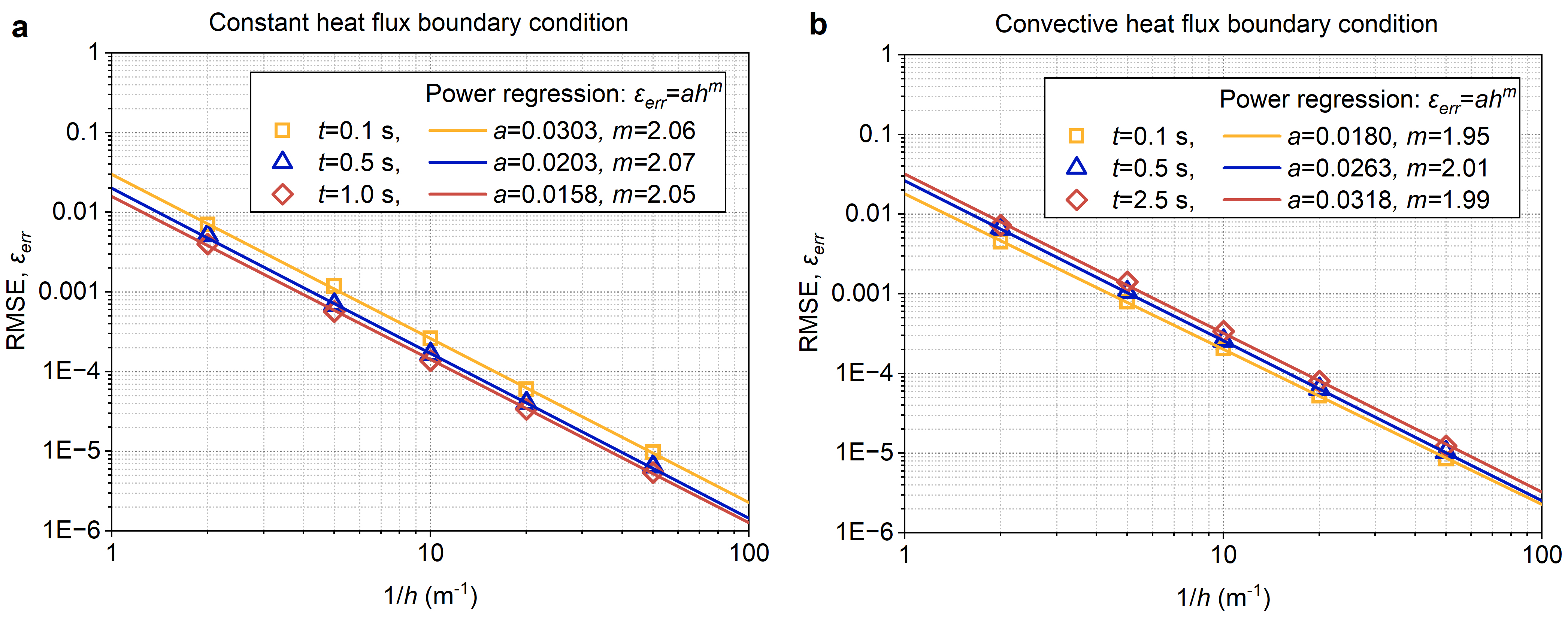}
    \caption{Transient heat transfer in a 1D semi-infinite Rod: Convergence test with mesh refinement for (a) constant heat flux boundary and (b) convective heat flux boundary.}
    \label{fig: ex1-3}
\end{figure}

To verify the convergence of the method, we further tested the error $\varepsilon_{\text{err}}$ under different mesh sizes. Five mesh sizes were considered: 0.5, 0.2, 0.1, 0.05, and 0.02 m. Since explicit integration is used, the time step must be sufficiently small to ensure stability for each case. As the time step size also influences the magnitude of the error, we ensured a fair comparison by setting the time step for each mesh size to $0.1t_{\text{CFL}}$. The corresponding time steps were $2.5 \times 10^{-2}$, $4 \times 10^{-3}$, $1 \times 10^{-3}$, $2.5 \times 10^{-4}$, and $4 \times 10^{-5}\ \mathrm{s}$, respectively. In this study, we tested only the nonconforming boundary condition. Fig.~\ref{fig: ex1-3} plots the variation of $\varepsilon_{\text{err}}$ against $1/h$ in log-log space. It is evident that for both constant and convective heat flux boundaries, the method achieves a clear second-order convergence rate.


\subsection{2D example: Heating of a circular ring}
\begin{figure}[!htb]
    \centering
    \includegraphics[width=0.8\linewidth]{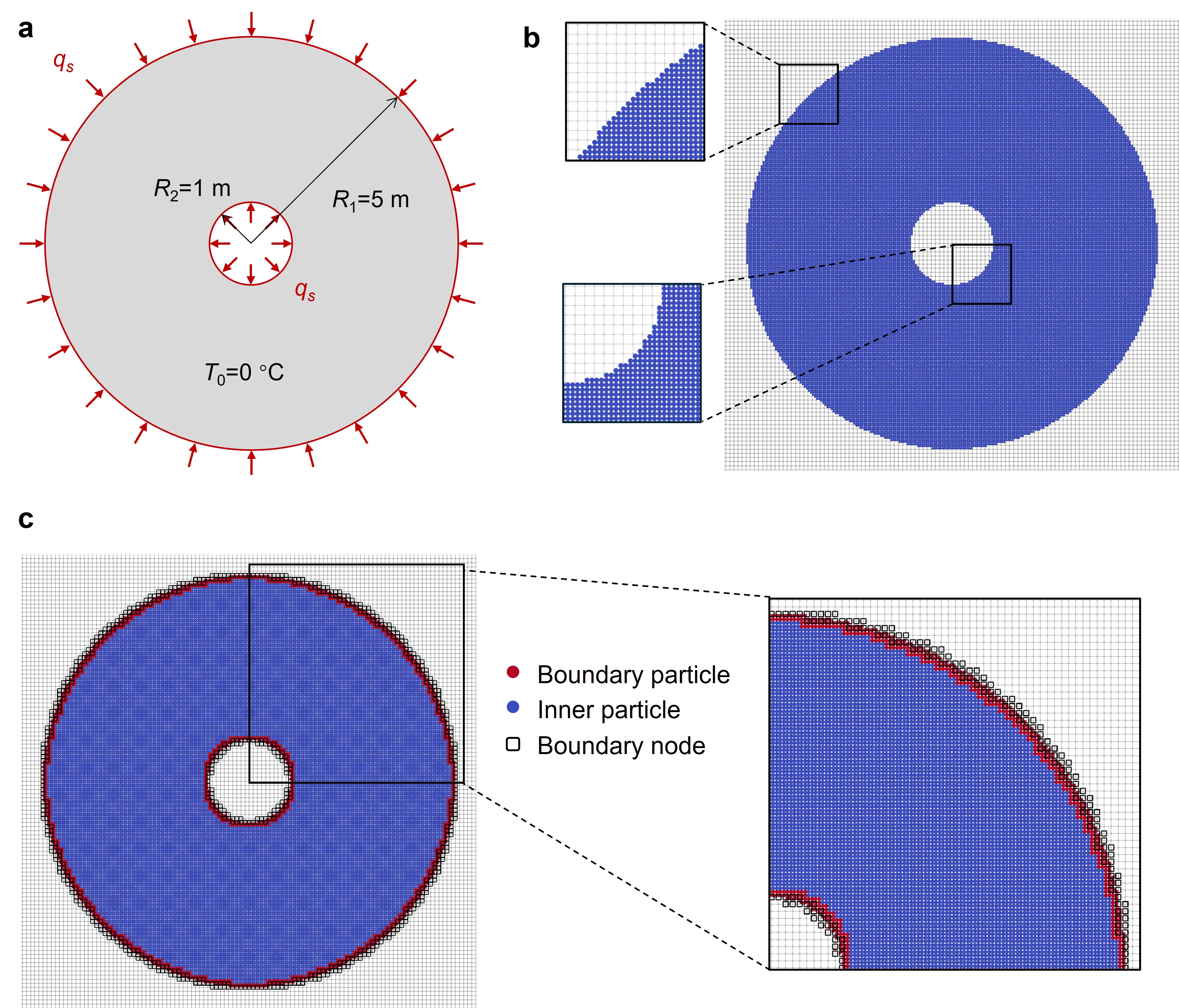}
    \caption{Bidirectional heating of circular ring: (a) model geometry, initial and boundary conditions, (b) background mesh and material point discretizations, and (c) detected surface nodes and surface particles based on volume fraction method.}
    \label{fig: ex2-1}
\end{figure}

The second example simulates the heating of a two-dimensional (2D) circular ring, aiming to verify the accuracy of VHFM for curved boundaries. As depicted in Fig.~\ref{fig: ex2-1}a, the outer and inner radii of the ring are $R_1 = 5\,\mathrm{m}$ and $R_2 = 1\,\mathrm{m}$, respectively. A heat flux $q_s$ is applied simultaneously to both the inner and outer boundaries of the circular ring. The initial temperature of the circular ring is set to $0\,^\circ\mathrm{C}$. Two types of heat flux boundary conditions are considered: a constant heat flux $q_s = 1\,\mathrm{W/m^2}$, and a convective heat flux boundary condition $q_s = \gamma(T - T_a)$, with the ambient temperature $T_a = 1\,^\circ\mathrm{C}$ and the heat transfer coefficient $\gamma = 1\,\mathrm{W/(m^2\cdot{^\circ}C)}$. Again, we assume unity material for the ring, i.e., density $\rho = 1\,\mathrm{kg/m^3}$, specific heat capacity $c = 1\,\mathrm{J/(kg\cdot{^\circ}C)}$, and thermal conductivity $\kappa = 1\,\mathrm{W/(m\cdot{^\circ}C)}$.

A quadrilateral background grid with a mesh size of $0.1\,\mathrm{m}$ is used, and the material point size is set to $0.05\,\mathrm{m}$. To discretize the circular ring, we first generate a $5\,\mathrm{m} \times 5\,\mathrm{m}$ square and then remove the material points outside the ring region. The resulting distribution of material points is shown in Fig.~\ref{fig: ex2-1}b. It can be observed that the circular ring boundary appears as a staircase pattern and does not align with the background grid. The time step is set to $\Delta t = 1\times10^{-3}$ s. The volume fraction threshold is set to 0.55, and the detected surface nodes and surface particles are shown in Fig.~\ref{fig: ex2-1}c. Since no available analytical solution for this problem, we compare the MPM result with FDM result. The FDM algorithm adopted for this 2D is shown in \ref{sec: FDM for ring}.

\begin{figure}[!htb]
    \centering
    \includegraphics[width=1\linewidth]{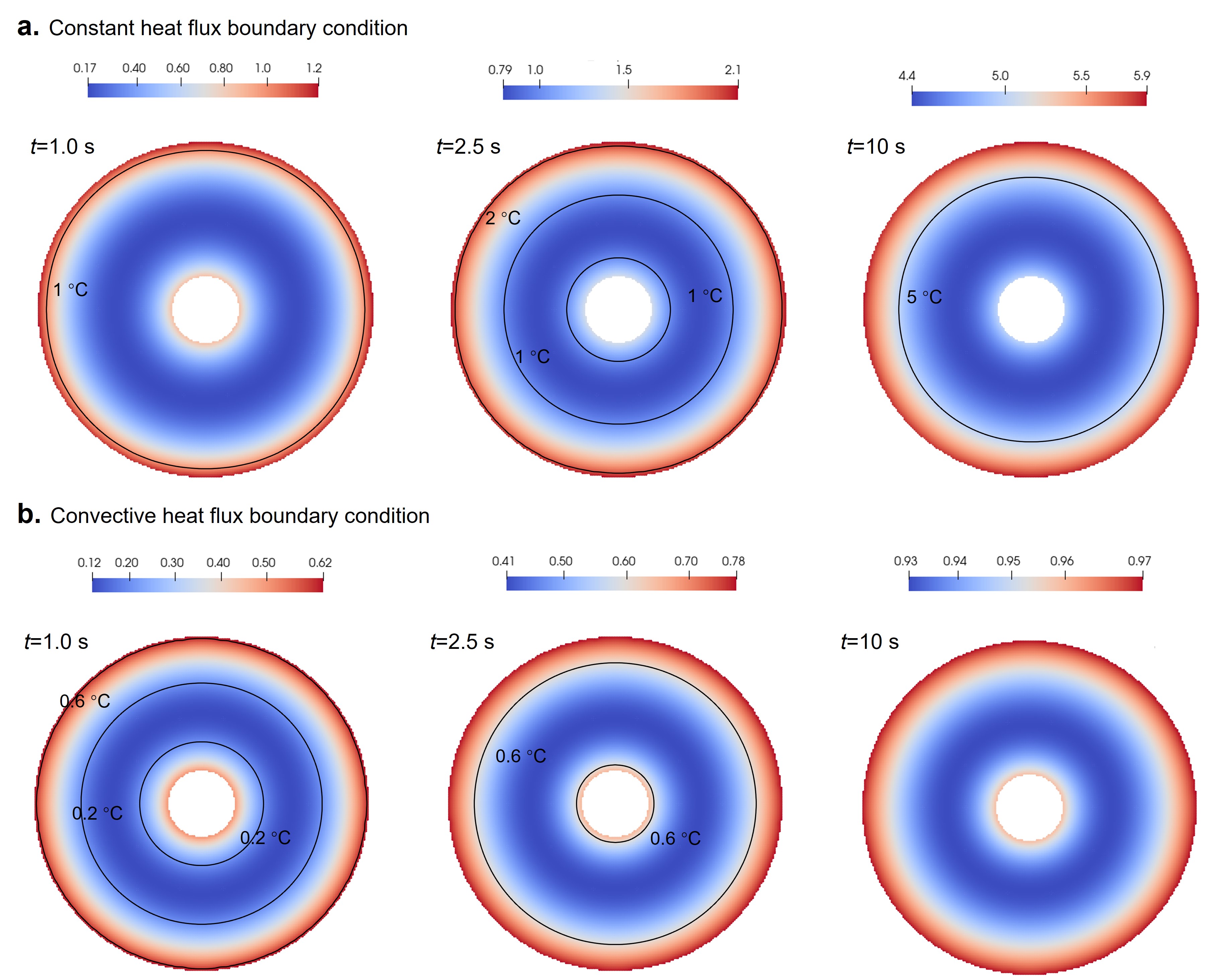}
    \caption{Bidirectional heating of a circular ring: temperature contours for (a) constant heat flux boundary condition and (b) convective heat flux boundary condition at $t = 1.0$, $2.5$, and $10\,\mathrm{s}$. The black lines show the isotherms.}
    \label{fig: ex2-2}
\end{figure}

\begin{figure}[!htb]
    \centering
    \includegraphics[width=1\linewidth]{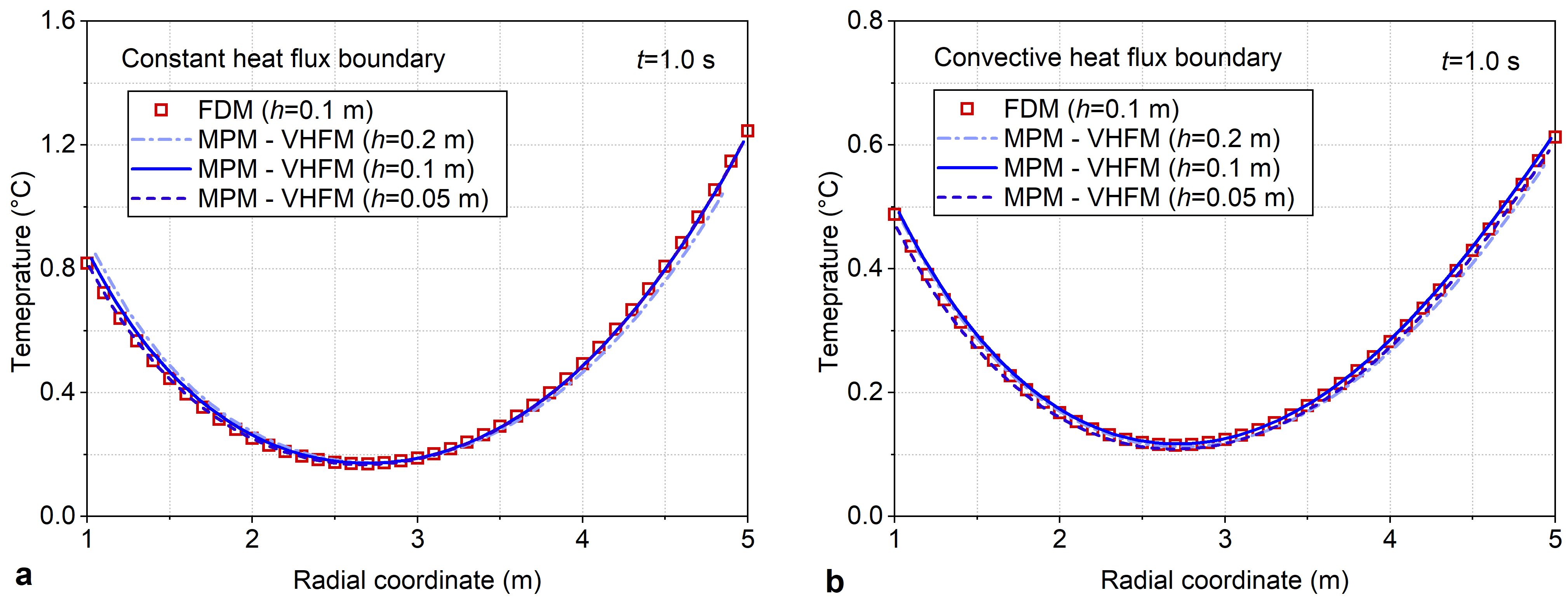}
    \caption{Bidirectional heating of a circular ring: comparison of radial temperature distributions at $t = 1.0$ s obtained using FDM and MPM with VHFM for different mesh sizes, under (a) constant heat flux boundary condition and (b) convective heat flux boundary condition .}
    \label{fig: ex2-3}
\end{figure}

\begin{figure}[!htb]
    \centering
    \includegraphics[width=1\linewidth]{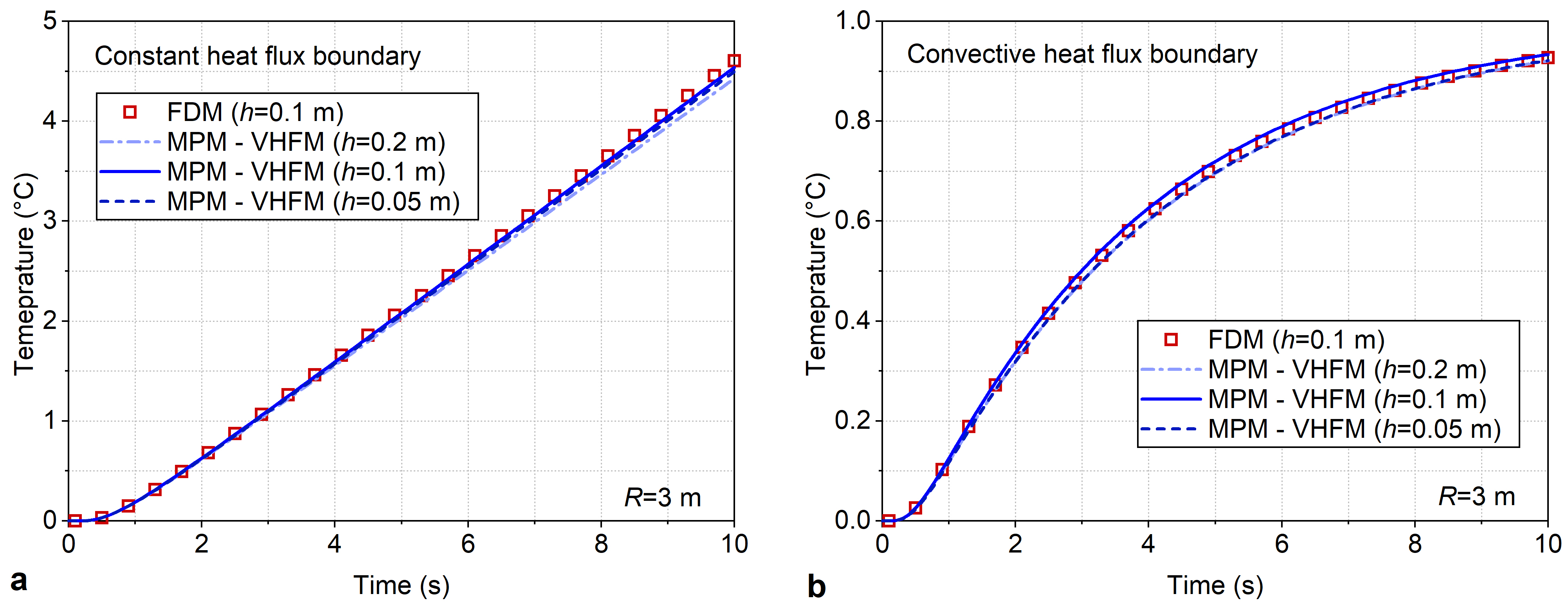}
    \caption{Bidirectional heating of a circular ring: comparison of temperature evolution at a radial distance of $R=$ 3 m, obtained using FDM and MPM with VHFM for different mesh sizes, under (a) constant heat flux boundary condition and (b) convective heat flux boundary condition.}
    \label{fig: ex2-4}
\end{figure}

\begin{figure}[!htb]
    \centering
    \includegraphics[width=1\linewidth]{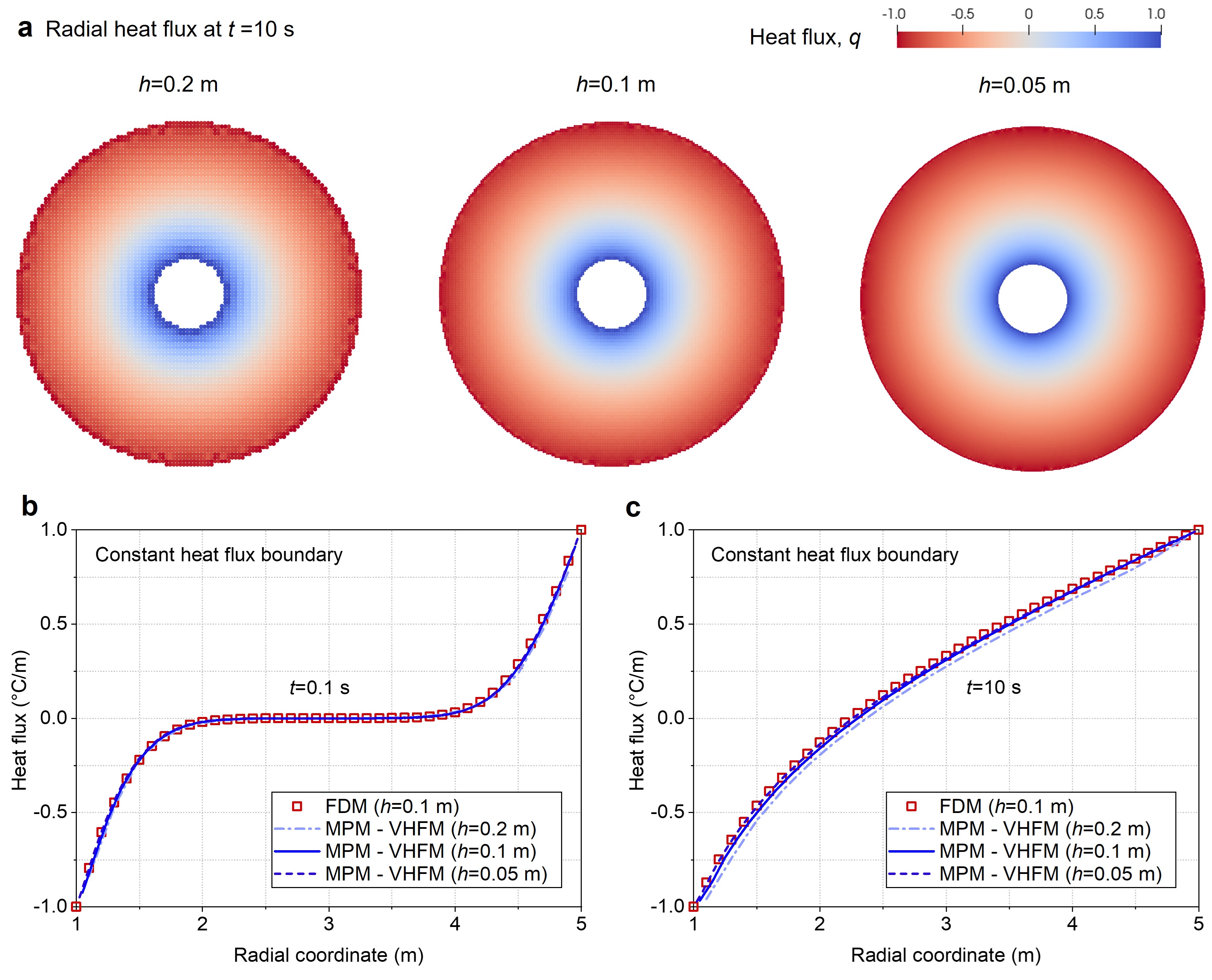}
    \caption{Bidirectional heating of a circular ring under constant heat flux boundary condition: (a) radial heat flux contours with different mesh sizes at $t=10$ s; (b) heat flux distribution along the radial direction at $t=0.1$ s; and (c) heat flux distribution along the radial direction at $t=1.0$ s.}
    \label{fig: ex2-5}
\end{figure}

Fig.~\ref{fig: ex2-3} quantitatively compares the radial temperature distribution of the ring at $t = 1\ \mathrm{s}$, obtained from MPM simulations and FDM calculations. For this problem, the FDM solution is based on a 1D axisymmetric model in polar coordinates, using the same mesh size as the MPM ($0.1\,\mathrm{m}$). The second-order central difference scheme and three-point one-sided difference for boundary heat flux imposition are employed in the FDM to ensure the accuracy. In the MPM simulations, in addition to the $0.1\,\mathrm{m}$ grid, two other grid sizes, $0.2\,\mathrm{m}$ and $0.05\,\mathrm{m}$, are also tested. The results indicate that the MPM and FDM solutions are in close agreement, and the predictions from MPM with VHFM show minimal variation across different mesh sizes. This demonstrates that even with relatively coarse grids, simulations based on VHFM can still produce reliable results.

Furthermore, Fig.~\ref{fig: ex2-4} compares the temperature evolution at a radial position of $3\,\mathrm{m}$ (i.e., along the central axis of the ring). Under the constant heat flux condition, the temperature increases continuously, while under the convective heat flux condition, the temperature gradually approaches the ambient temperature of $1\,^\circ\mathrm{C}$. In both cases, the temperature evolution predicted by MPM aligns closely with the FDM results, with simulations across different grid sizes yielding nearly identical results. This further validates the accuracy and robustness of the proposed VHFM in handling non-aligned thermal boundary conditions.

Fig.~\ref{fig: ex2-5}a compares the radial heat flux distribution under the constant heat flux condition after reaching steady state ($t = 10\,\mathrm{s}$) for different mesh sizes. The results show that the heat flux values at the inner and outer boundaries are $-1\,^\circ\mathrm{C/m}$ (directed toward the center) and $1\,^\circ\mathrm{C/m}$ (directed away from the center), respectively, which are perfectly consistent with the imposed boundary conditions. Even for a coarse mesh ($h = 0.2\,\mathrm{m}$), the differences in the heat flux distribution compared to finer meshes are negligible. Figs.~\ref{fig: ex2-5}b and c further present a quantitative comparison of the radial heat flux at $t = 0.1\,\mathrm{s}$ and $t = 10\,\mathrm{s}$, including results from FDM and MPM simulations with three different mesh sizes. The results demonstrate that the MPM simulations based on VHFM are in excellent agreement with the FDM solution, both in the early heating stage ($t = 0.1\,\mathrm{s}$ ) and at the final steady state ($t = 10\,\mathrm{s}$). Moreover, the results across different mesh sizes exhibit good consistency. These findings further validate the accuracy of the proposed VHFM method for thermal boundary imposition.

\subsection{3D example: Cooling of a sphere}
\begin{figure}[!htb]
    \centering
    \includegraphics[width=0.9\linewidth]{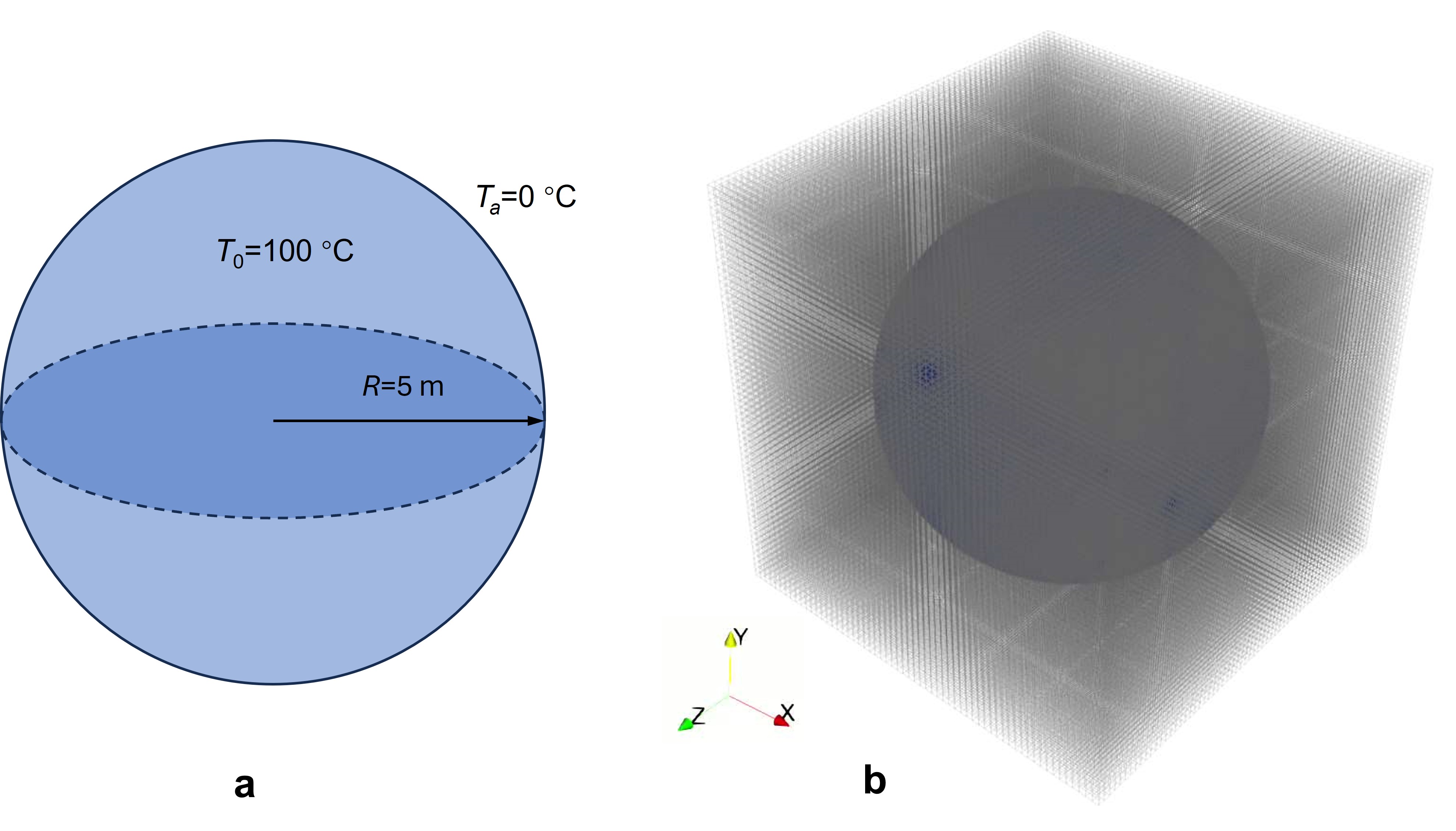}
    \caption{Cooling of a sphere: (a) geometry, initial and boundary conditions; (b) background mesh and material discretization. }
    \label{fig: ex5-1}
\end{figure}

The third numerical example simulates the transient cooling of a three-dimensional (3D) sphere. The sphere has a radius of \SI{5}{\meter}, a uniform initial temperature of \SI{100}{\degreeCelsius}, and is exposed to an ambient temperature of \SI{0}{\degreeCelsius} with a convective heat transfer coefficient of $\gamma = 1\,\mathrm{W/(m^2\cdot{^\circ}C)}$. Unit material properties were employed. The simulation utilized a uniform hexahedral background mesh with an element size of \SI{0.2}{\meter} (see Fig.~\ref{fig: ex5-1}b). Two distinct particle-per-cell (PPC) configurations are examined: 8 particles and 27 particles per background cell. The time step is set to $\Delta t = 1 \times 10^{-2}\,\text{s}$, and the volume fraction threshold $\eta$ is consistently maintained at 0.55. Since no available closed-form solution exists for this problem, validation is performed against a finite difference method (FDM) reference solution. The problem can be reduced to 1D heat transfer in spherical coordinates; the specific FDM algorithm is detailed in~\ref{app: 3D sphere}.

\begin{figure}[!htb]
    \centering
    \includegraphics[width=0.95\linewidth]{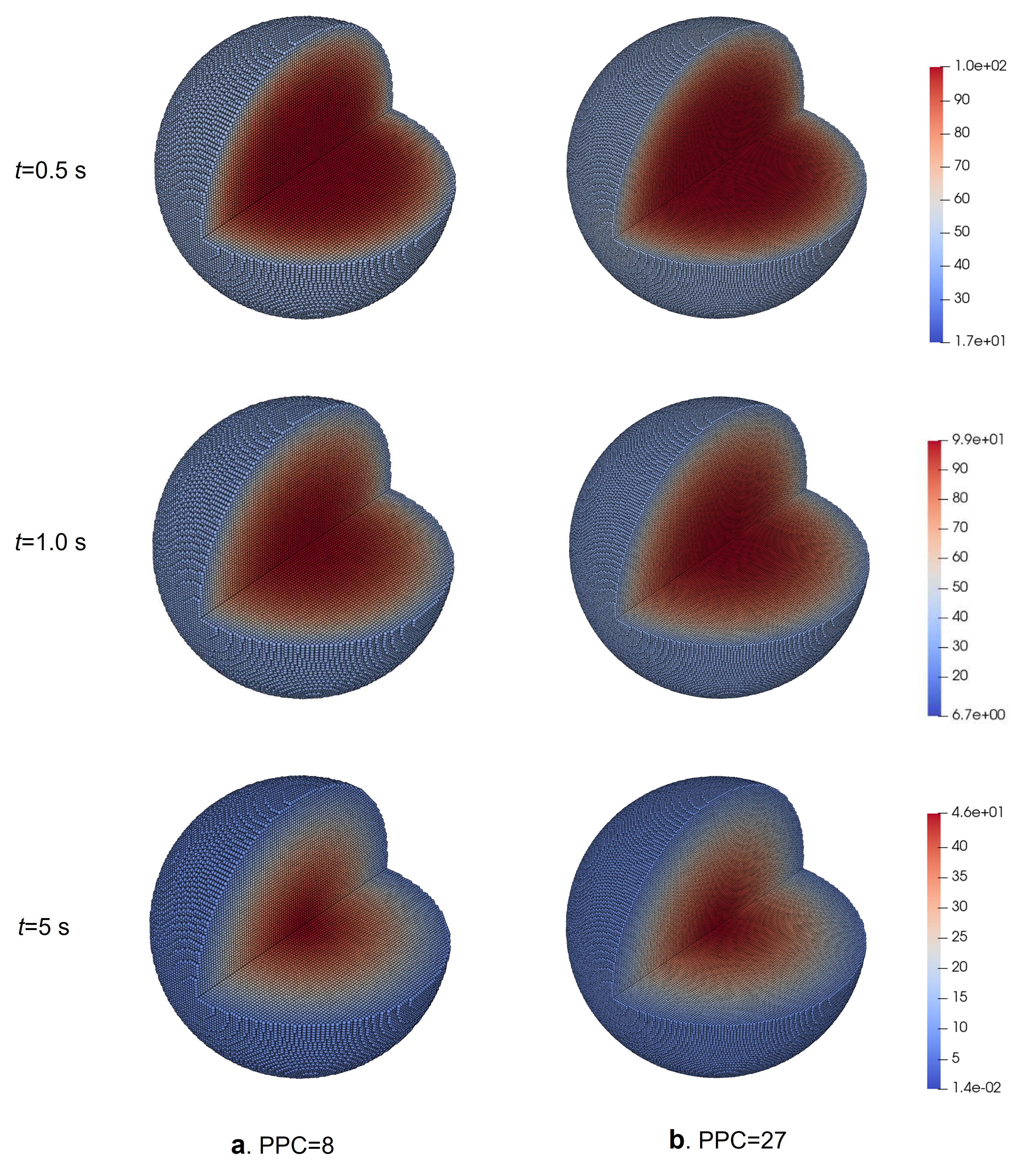}
    \caption{Cooling of a sphere: temperature distribution at various time instants for the case (a) PPC =8 and (b) PPC=27. }
    \label{fig: ex5-2}
\end{figure}

\begin{figure}[!htb]
    \centering
    \includegraphics[width=0.55\linewidth]{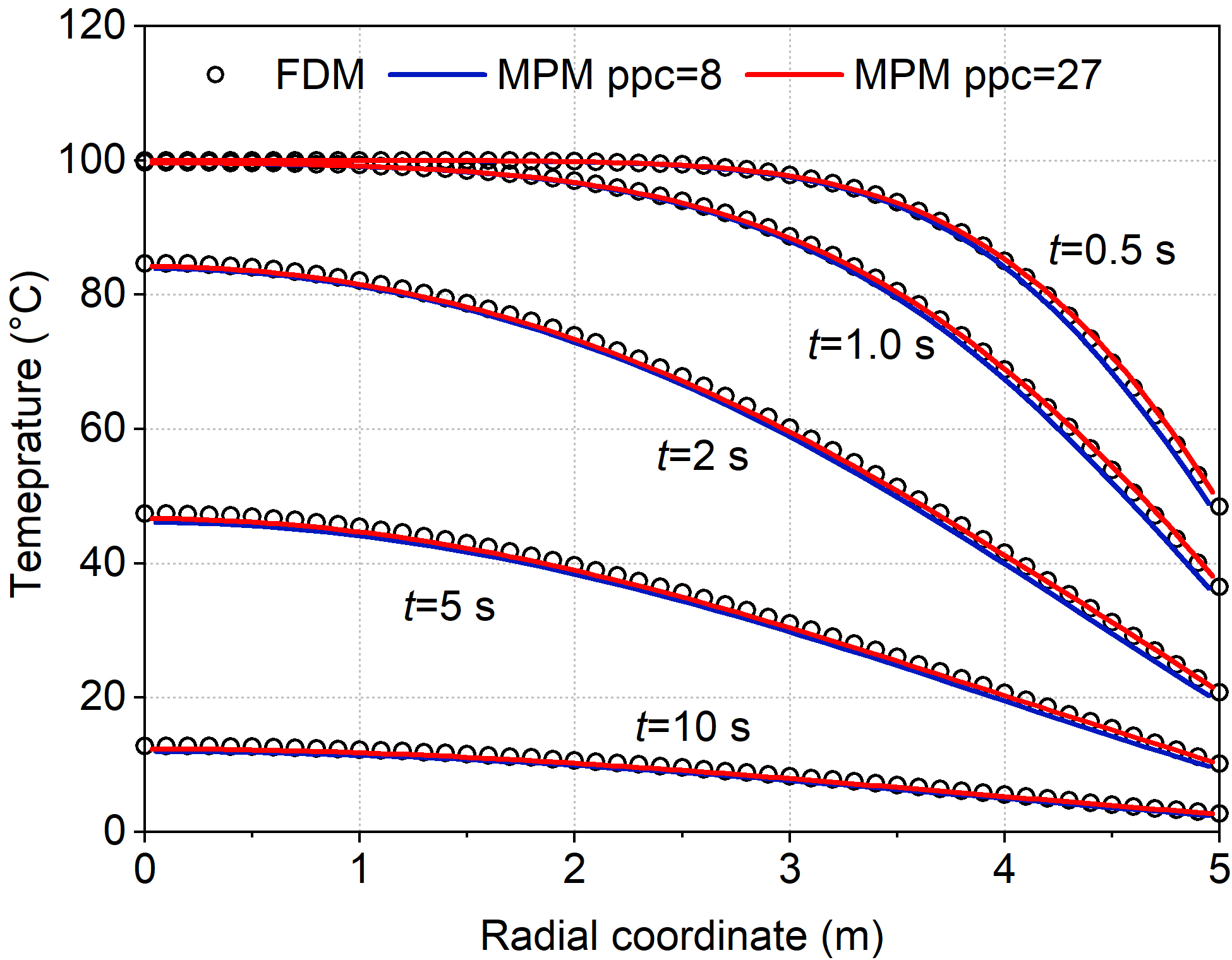}
    \caption{Cooling of a sphere: comparison of temperature along the radial direction solved by MPM and FDM at $t=$ 0.5, 1.0, 2.0, 5.0, and 10.0 s}
    \label{fig: ex5-3}
\end{figure}

Fig.~\ref{fig: ex5-2} illustrates the temperature distribution within the sphere at various time instances, demonstrating the progression of cooling from the surface toward the center. The results for both PPC configurations are visually nearly identical. A quantitative comparison between the MPM predictions and the FDM reference for the radial temperature profile at selected times is presented in Fig.~\ref{fig: ex5-3}. The case with PPC = 8 shows good agreement with the FDM results, while the case with PPC = 27 exhibits even closer agreement, confirming that increasing the number of particles per cell improves numerical accuracy. The FDM solution is considered the more accurate benchmark here, as it solves a simplified 1D problem and allows for more precise imposition of boundary conditions. However, FDM is generally restricted to problems with simple, regular geometries; for complex shapes, particle-based methods offer superior flexibility. This example demonstrates that the proposed Volume Fraction Method (VHFM) provides accurate and robust simulations for 3D transient heat transfer problems.

\subsection{Moving boundary: Heating of a square block}
\begin{figure}[!htb]
    \centering
    \includegraphics[width=1\linewidth]{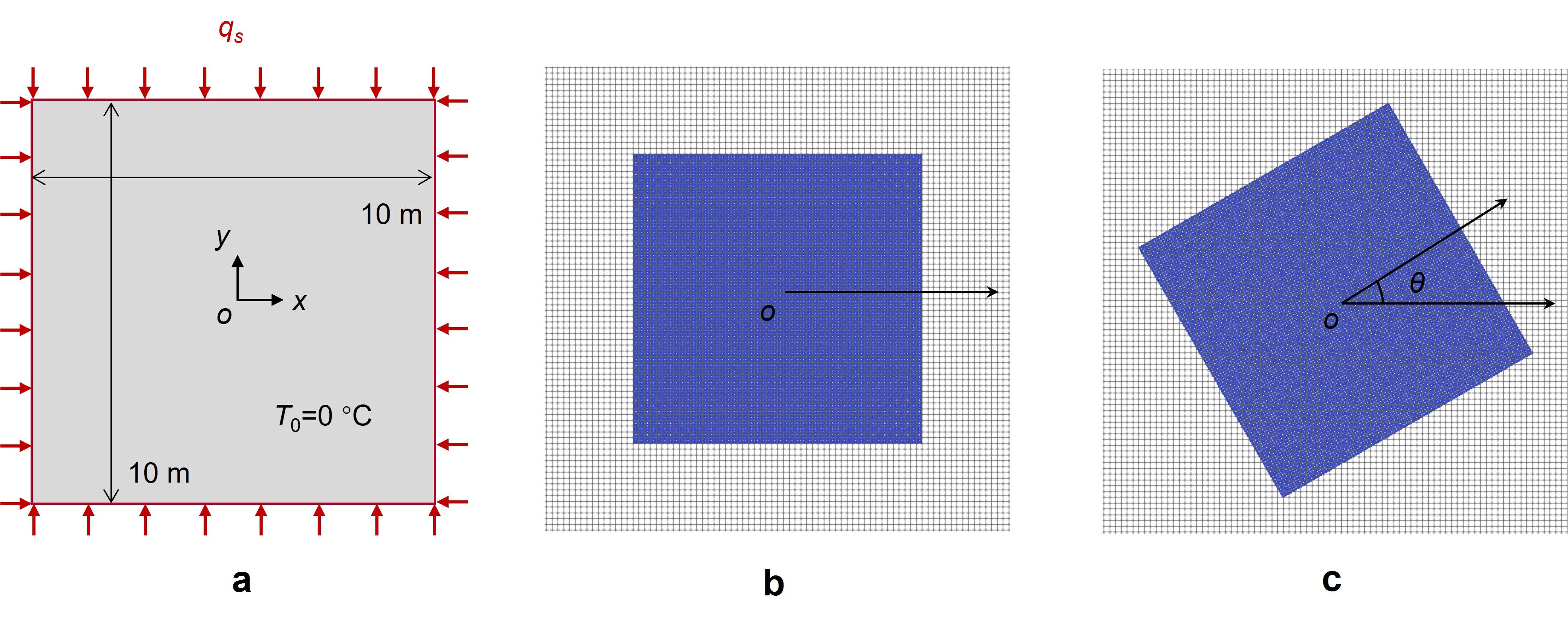}
    \caption{Heating of a rotating square block: (a) geometry, initial and boundary conditions, (b) conforming boundary condition, and (c) nonconforming boundary condition.}
    \label{fig: ex3-1}
\end{figure}

In the previous example, although the boundary is circular, its geometric configuration remains fixed. To verify that the proposed VHFM can accurately impose thermal boundary conditions even under dynamically changing boundary configurations, the third example considers a 2D square domain. The initial boundary of the square is aligned with the grid, after which it is rotated by a certain angle. Convective heat flux boundary conditions are applied simultaneously to all four sides of the square, and only the convective heat flux case is presented here. The model geometry, initial and boundary conditions are depicted in Fig.~\ref{fig: ex3-1}a. The side length of the square is $5\,\mathrm{m}$; the initial temperature is $0\,^\circ\mathrm{C}$; the ambient temperature is $1\,^\circ\mathrm{C}$; and the convective heat transfer coefficient is $1\,\mathrm{W/(m^2\cdot {^\circ}C)}$. The background grid size is $0.2\,\mathrm{m}$, and four material points are assigned within each grid cell. Similar to the previous case, the material properties are set to unit values. The simulation time step size is set to $1 \times 10^{-2}\,\mathrm{s}$.

In the MPM simulations, we consider the following two scenarios: (1) fixed rotation angles of $15^\circ$, $30^\circ$, and $45^\circ$; and (2) fixed rotational speeds (denoted by $\omega$) of $1/16$, $1/4$, 1, and 4 revolutions per second (unit: r/s). The conforming case, where the boundary is aligned with the grid, is used as a reference (Fig.~\ref{fig: ex3-1}b). Based on the findings from the first example in Section~\ref{sec: Heating of 1D semi-infinite Rod}, applying boundary conditions using VHFM produces results equivalent to directly imposing them at the nodes, both of which are more accurate than applying them at the material points. Therefore, VHFM is directly used for boundary condition imposition in the reference case. Since no analytical solution exists for this 2D problem, the FDM results are used as a reference for comparison. Details of the FDM algorithm for this problem can be found in~\ref{sec: FDM for 2D square}.

\begin{figure}
    \centering
    \includegraphics[width=1\linewidth]{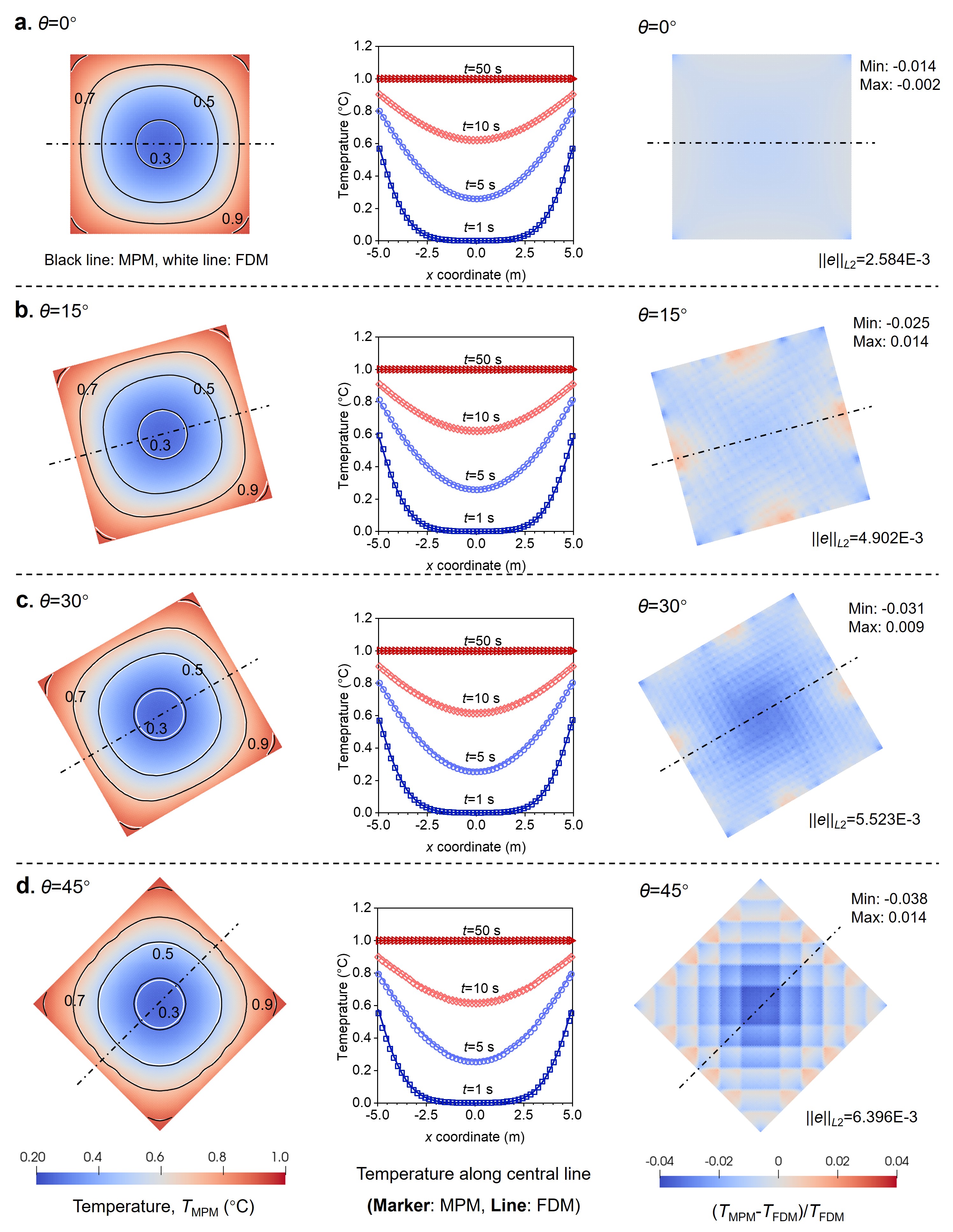}
    \caption{Heating of a rotating square block: simulation results for different rotation angles: (a) $\theta = 0^\circ$, (b) $\theta = 15^\circ$, (c) $\theta = 30^\circ$, and (d) $\theta = 45^\circ$. The first column shows the temperature contours simulated by MPM with VHFM at  $t = 5\,\mathrm{s}$, where the black and white isothermal lines represent MPM and FDM results, respectively. The second column compares the temperature distribution along the central line simulated by MPM and FDM at $t = 1\,\mathrm{s}$, $5\,\mathrm{s}$, $10\,\mathrm{s}$, and $50\,\mathrm{s}$. The third column presents the relative error of the MPM results compared to the FDM results at $t = 5\,\mathrm{s}$.}
    \label{fig: ex3-2}
\end{figure}

Figs. \ref{fig: ex3-2}a-d present the simulation results for cases with different rotation angles $\theta$. For each case, we first show the temperature contour plot computed using the MPM at $t = 5$ s. Superimposed on these contours are the temperature iso-lines ($0.3$, $0.5$, $0.7$, and $0.9$ $^\circ\mathrm{C}$) calculated by both MPM and the FDM for direct comparison. Additionally, the temporal evolution of the temperature profiles along the horizontal central axis of the square, obtained from both the MPM and FDM at $t = 1, 5, 10,$ and $50$ s, is compared. Moreover, the relative error between the two methods, defined as $(T_{\text{MPM}} - T_{\text{FDM}})/T_{\text{FDM}}$, is analyzed. The overall accuracy is quantified by computing the L2 norm of the relative error across all material points,
\begin{equation}
    \|e\|_{L^2} = \sqrt{\frac{1}{N_p} \sum_{p=1}^{N_p} (T_{\text{MPM}} - T_{\text{FDM}})_p^2}\,,
    \label{eq:error_norm}
\end{equation}
where $N_p$ denotes the total number of material points.

\begin{figure}[!htb]
    \centering
    \includegraphics[width=0.65\linewidth]{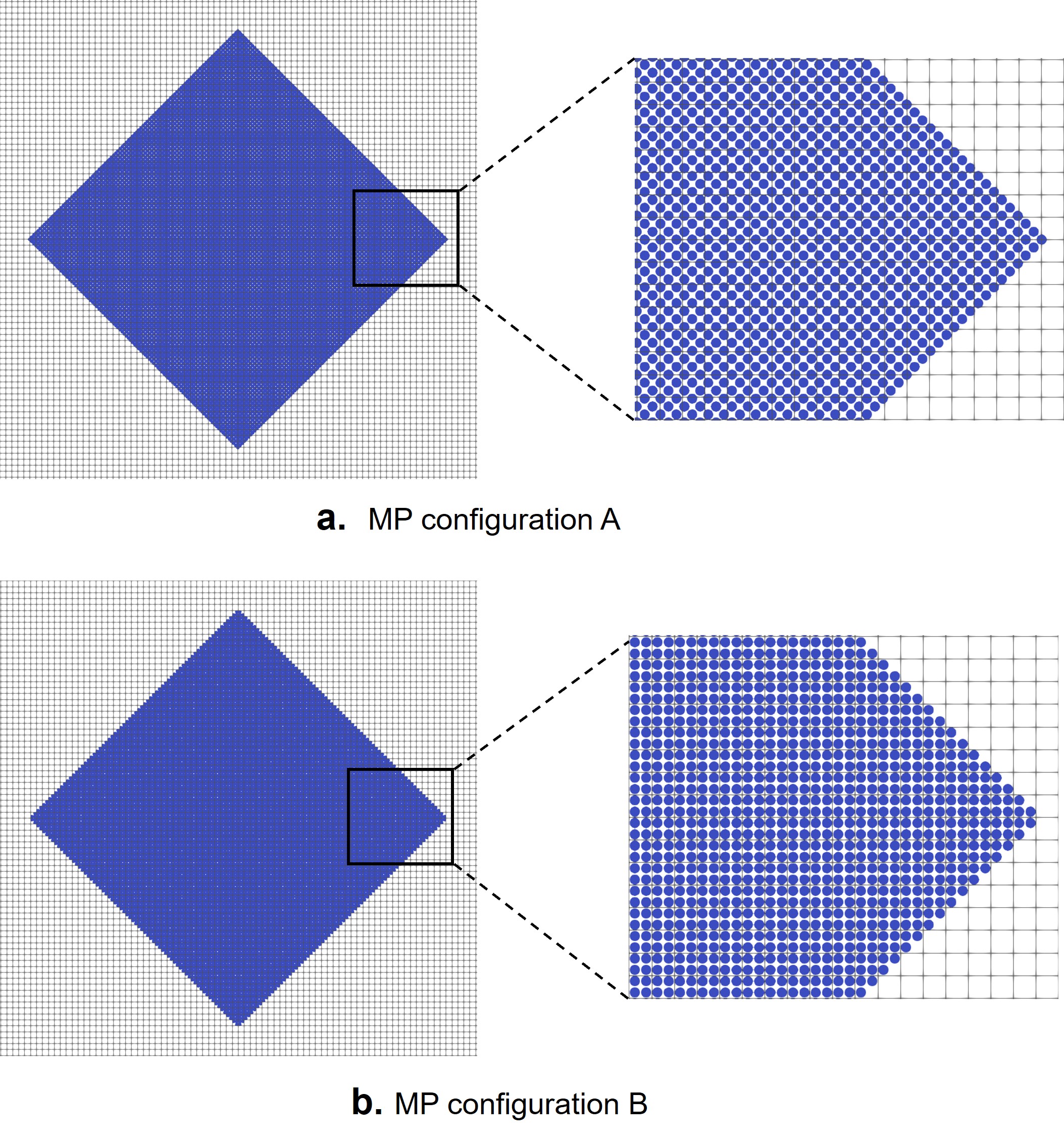}
    \caption{Heating of a rotating square block: (a) MP Configuration A - suboptimal MP distribution; (b) MP Configuration B - optimal MP distribution.}
    \label{fig: ex3-3}
\end{figure}

The results, as shown in Fig.~\ref{fig: ex3-2}a, demonstrate excellent agreement between the MPM and the FDM for the case of $\theta = 0^\circ$ (i.e., conforming boundary configuration), with a very small error norm $\|e\|_{L^2}$. As the rotation angle increases, the MPM and FDM results remain in good overall agreement. However, a slight but consistent increase in $\|e\|_{L^2}$ is observed with larger $\theta$. Notably, for $\theta = 45^\circ$, the temperature iso-lines exhibit some spatial oscillations, and the spatial distribution of the relative error exhibits a distinct checkerboard pattern. This pattern is attributed to the suboptimal distribution of material points relative to the background grid after a $45^\circ$ rotation. As illustrated in Fig.~\ref{fig: ex3-3}a, the rotated material points are distributed irregularly with respect to the computational grid. Most of these points are displaced from optimized integration points, the Gaussian points. It is also clear that some of the entire rows or columns of material points coincide with grid lines - positions where the integration accuracy is minimal. This suboptimal discretization is the primary cause of the checkerboard pattern observed in the error contour plot of Fig. \ref{fig: ex3-3}d.

\begin{figure}[!htb]
    \centering
    \includegraphics[width=1\linewidth]{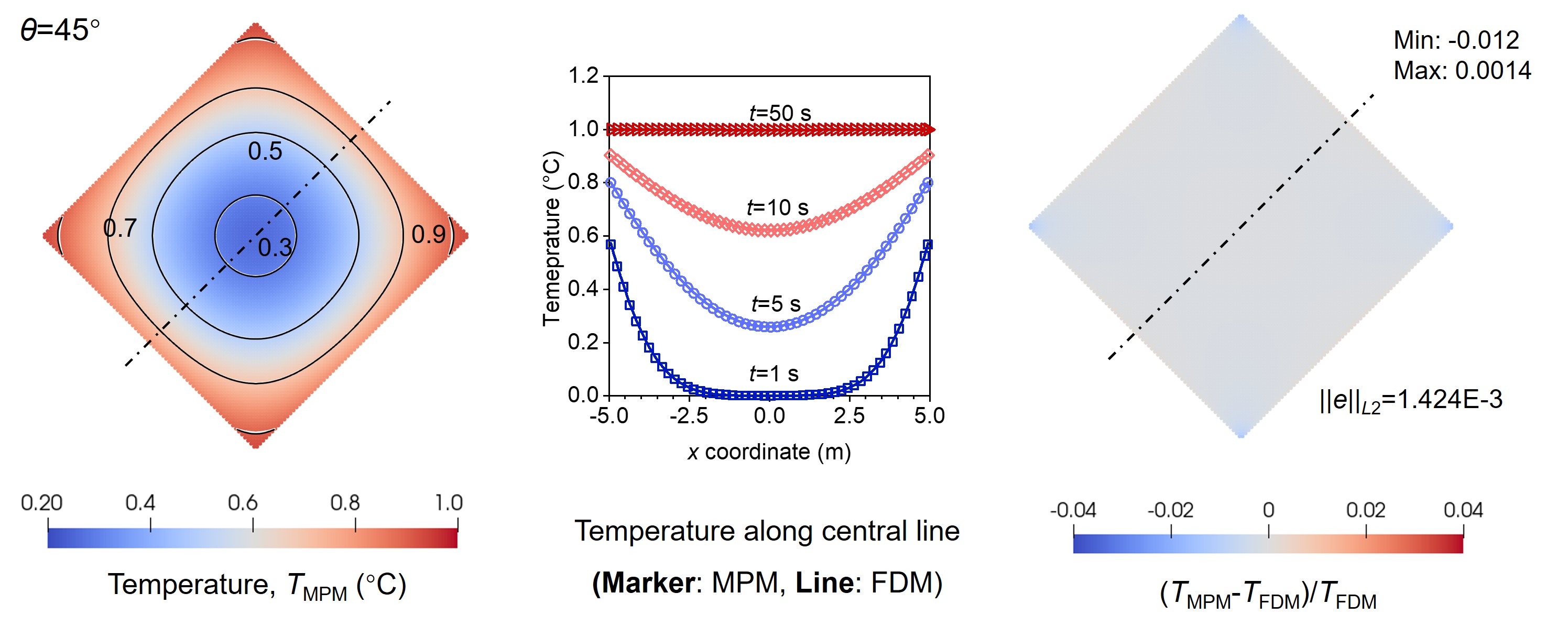}
    \caption{Heating of a rotating square block: simulation results for $\theta = 45^\circ$ using MP Configuration B.}
    \label{fig: ex3-4}
\end{figure}

Therefore, the increased error for larger rotation angles likely stems predominantly from the material points deviating from optimal integration points, rather than from inaccuracies inherent to the boundary condition imposition by the VHFM. To verify this, we simulated the $\theta = 45^\circ$ case using a different, optimized material point configuration (referred to as MP Configuration B), shown in Fig. \ref{fig: ex3-3}b. This configuration was generated by first creating a regular distribution of material points (PPC=4) across the entire background mesh and then removing points located in the region defined by $|x| + |y| > 5\sqrt{2}$. Thus, MP Configuration B models the same physical domain as Configuration A (Fig. \ref{fig: ex3-3}a) but with a more favorable material point distribution. The simulation results using Configuration B, presented in Fig. \ref{fig: ex3-4}, show an error distribution and magnitude similar to those of the conforming case ($\theta = 0^\circ$) in Fig. \ref{fig: ex3-2}a. This confirms that with an optimized internal discretization, nonconforming boundaries can yield accuracy comparable to conforming ones. It is important to note that even the original MP Configuration A maintains an error of the same order of magnitude as the conforming boundary case.

\begin{figure}[!htb]
    \centering
    \includegraphics[width=1\linewidth]{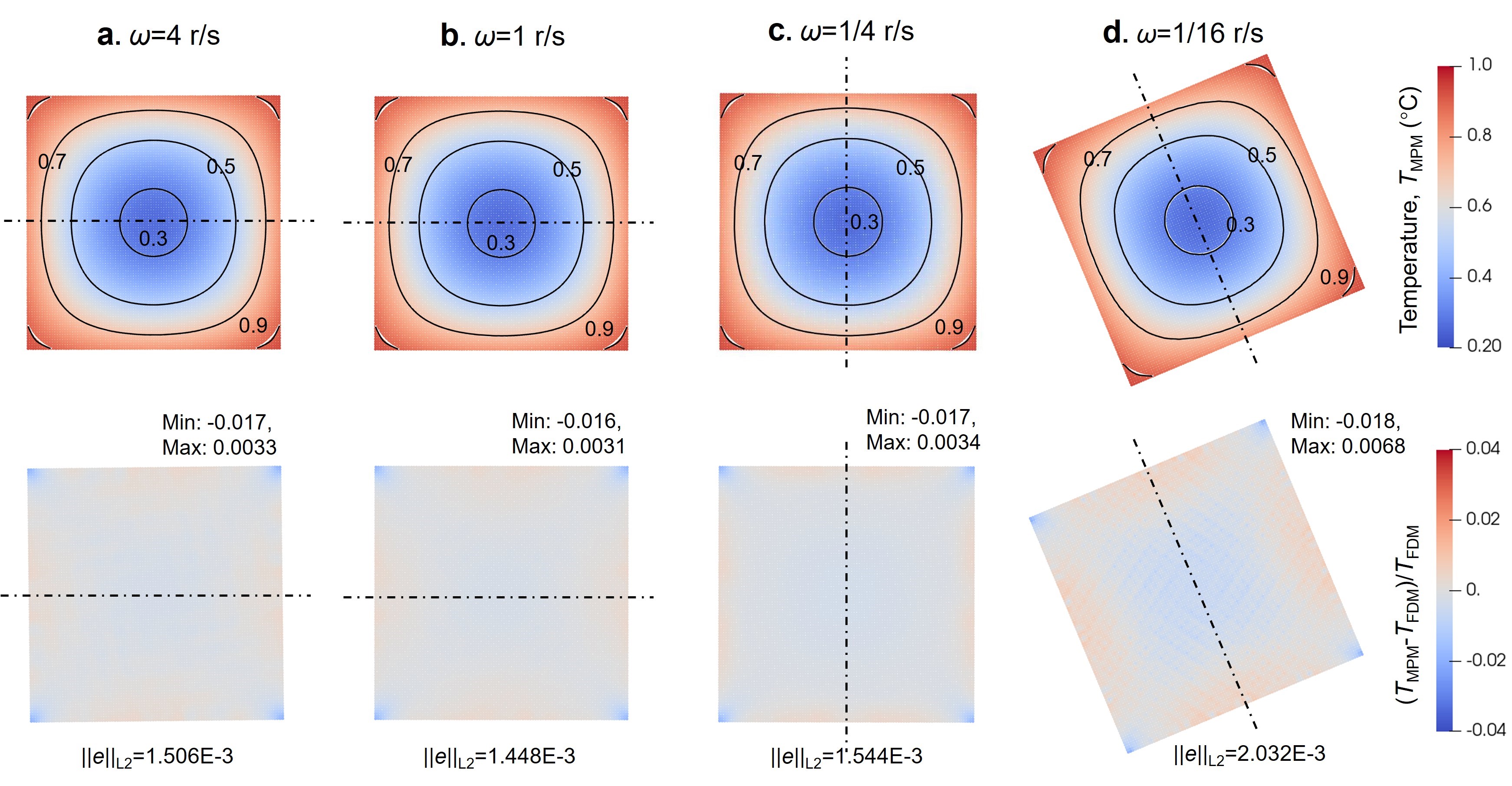}
    \caption{Heating of a rotating square block: temperature and relative error contours with different resolution rate: (a) $\omega=4\ \rm r/s$, (b) $\omega=1\ \rm r/s$, (c) $\omega=1/4\ \rm r/s$, and (d) $\omega=1/16\ \rm r/s$.}
    \label{fig: ex3-5}
\end{figure}

The above analysis pertains to Scenario A, which involves fixed, nonconforming boundaries. We now present results for Scenario B, where heat transfer occurs concurrently with the rigid-body rotation of the square. Fig. \ref{fig: ex3-5} compares the temperature contours and relative error distributions (at $t = 5$ s) for four different rotational velocities. The results show that the errors remain very small in all cases, with $\|e\|_{L^2}$ norms on par with those of the static conforming boundary case. This demonstrates that the proposed method remains highly accurate for simulating convective heat transfer even under dynamically evolving boundary conditions.

\subsection{Moving boundary: Cooling a rotating fan}
\begin{figure}[!htb]
    \centering
    \includegraphics[width=0.8\linewidth]{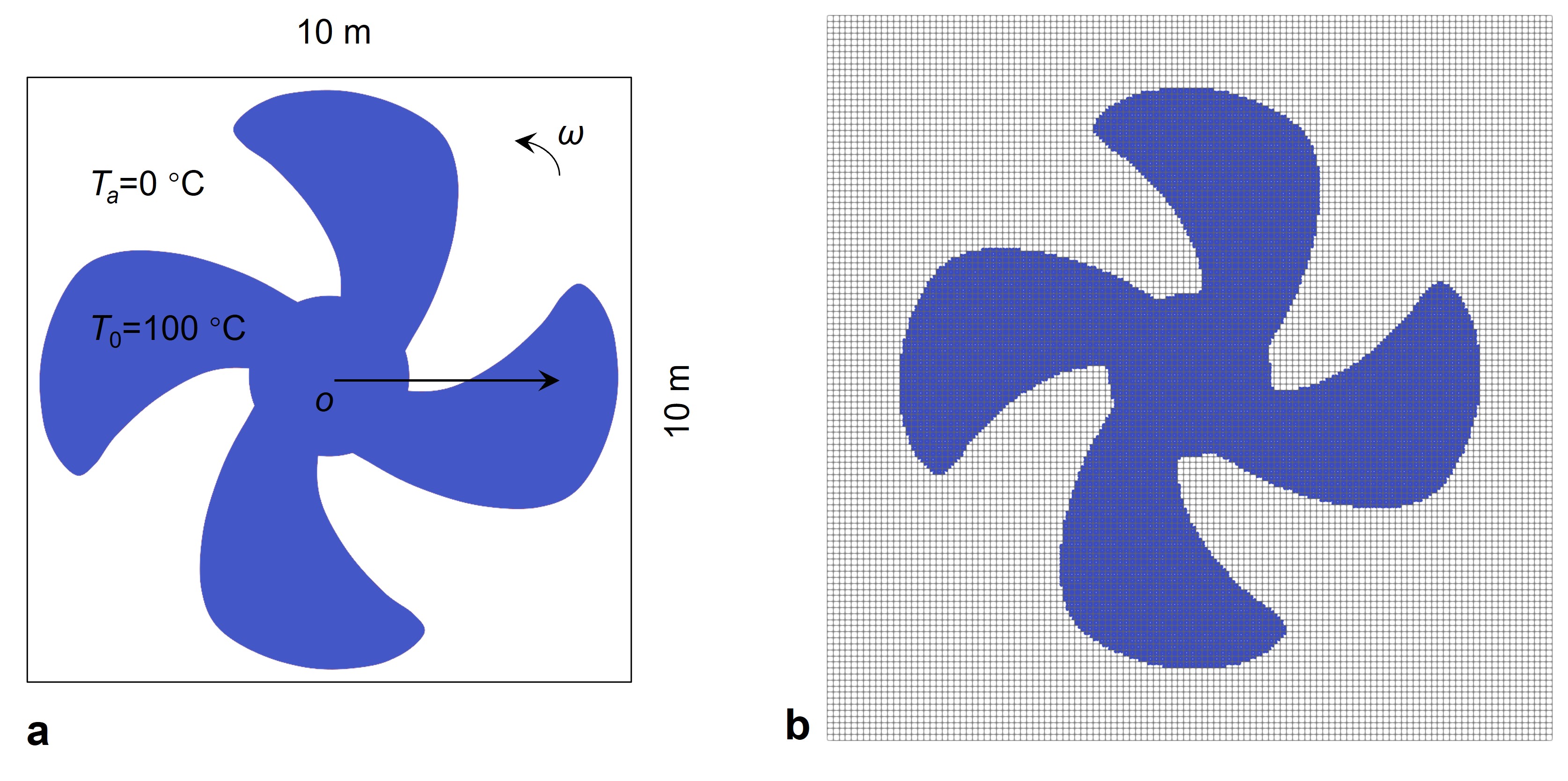}
    \caption{Cooling of a rotating fan: (a) geometry of the fan and (b) material discretizations.}
    \label{fig: ex4-1}
\end{figure}

In the final example, we simulate a cooling problem of a rotating fan. The geometry of the fan is shown in Fig.~\ref{fig: ex4-1}a, and its topological relationship with the background grid is illustrated in Fig.~\ref{fig: ex4-1}b. The initial temperature of the fan is set to $T_0 = 100^\circ\mathrm{C}$, while the ambient temperature is $T_a = 0^\circ\mathrm{C}$. The convective heat transfer coefficient is prescribed as $\gamma = 1\ \mathrm{W/(m^2\cdot {^\circ}C)}$. 
The fan rotates about its center with a constant angular velocity of $\omega = 1~\mathrm{r/s}$. The material point coordinates are generated directly from the absolute pixel coordinates of the input image. A background grid with a cell size of $h = 0.1~\mathrm{m}$ is first employed, with PPC=4. The time step is chosen as $\Delta t = 5 \times 10^{-3}~\mathrm{s}$.

\begin{figure}[!htb]
    \centering
    \includegraphics[width=0.9\linewidth]{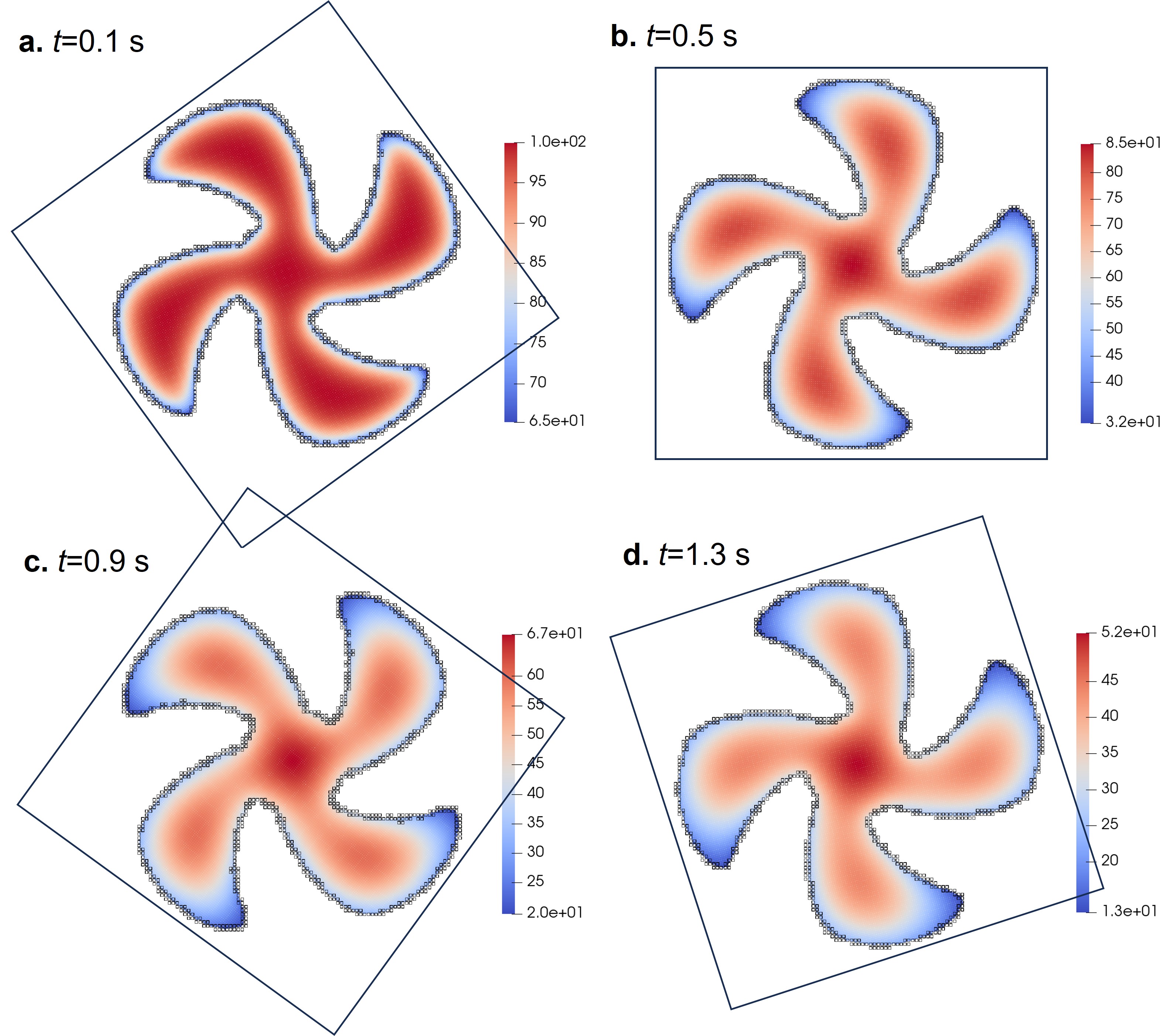}
    \caption{Cooling of a rotating fan: simulation results at different time instances.}
    \label{fig: ex4-2}
\end{figure}

\begin{figure}[!htb]
    \centering
    \includegraphics[width=0.95\linewidth]{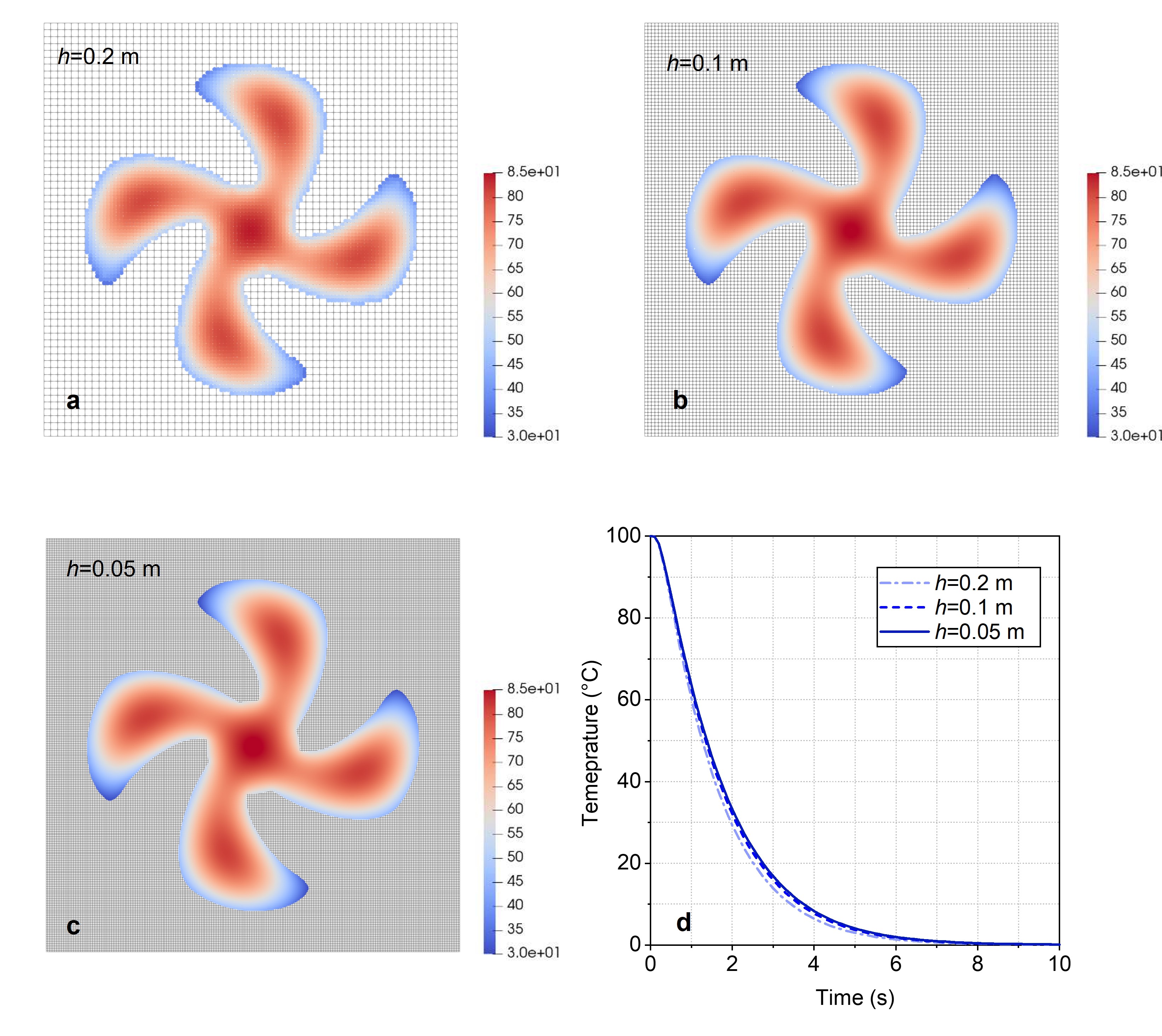}
    \caption{Cooling of a rotating fan: Mesh sensitivity analysis. (a–c) Temperature distribution at $t = 0.5\,\mathrm{s}$ for mesh sizes $h = 0.1\,\mathrm{m}$, $h = 0.05\,\mathrm{m}$, and $h = 0.025\,\mathrm{m}$, respectively; (d) Temperature evolution at the center of the fan for different mesh sizes.}
    \label{fig: ex4-3}
\end{figure}

Fig.~\ref{fig: ex4-2} presents the temperature contours at different time instants together with the detected surface nodes.  The results exhibit a smooth temperature distribution, and the surface node detection remains reasonable and stable throughout the rotational process. Since no analytical solution is available for validation, additional simulations are conducted using a coarser grid ($h = 0.2~\mathrm{m}$) and a finer grid ($h = 0.05~\mathrm{m}$) to examine solution consistency and convergence.  The corresponding time steps are set to $\Delta t = 2 \times 10^{-2}~\mathrm{s}$ and $\Delta t = 1.25 \times 10^{-3}~\mathrm{s}$, respectively. Figs.~\ref{fig: ex4-3}(a--c) compare the temperature distributions at $t = 0.5~\mathrm{s}$ for the three mesh resolutions. Good agreement is observed in both the overall temperature patterns and magnitudes. In addition, the temperature evolution at the center of the fan is examined, as shown in Fig.~\ref{fig: ex4-3}d. The temperature decreases from $100^\circ\mathrm{C}$ to $0^\circ\mathrm{C}$, and as the mesh is refined, the temperature histories converge to nearly a single curve.

This example demonstrates that the proposed VHFM is capable of accurately and robustly handling complex geometries and evolving boundary configurations.

\section{Conclusions}
\label{sec: Conclusions}
This study has presented a novel Virtual Heat Flux Method (VHFM) for the simple, accurate, and robust imposition of Neumann-type thermal boundary conditions within the Material Point Method. The method fundamentally addresses the challenge of nonconforming boundaries by recasting the boundary flux integral into a volumetric term via a conceptually simple virtual flux field. 

The proposed VHFM offers several key advantages: (1) It removes the necessity for explicit boundary tracking, surface reconstruction, or the use of specialized boundary particles, preserving the inherent efficiency of the standard MPM algorithm. (2) VHFM achieves accuracy on par with direct nodal imposition in conforming cases and delivers superior, consistent results for nonconforming, curved, and moving boundaries where conventional particle-based methods fail. (3) The formulation provides a unified framework readily extensible to other Neumann conditions (e.g., traction, fluid flux) through appropriate construction of the virtual field. (4) Our numerical experiments, spanning 1D to 3D, from simple geometries to a complex rotating fan, demonstrate the method's stability and second-order spatial convergence under large deformations, rotations, and complex boundary evolution.

The performance of VHFM is primarily governed by the quality of the internal material point discretization rather than the boundary non-conformity itself, as evidenced by the rotating square example. This underscores its practicality for real-world simulations where boundaries are rarely mesh-aligned. The VHFM framework opens promising avenues for future research. Immediate extensions include its integration into fully coupled thermo-hydro-mechanical-chemical (THMC) MPM formulations, where consistent flux boundary conditions are critical. Applying VHFM to problems involving phase change (e.g., melting/solidification) and frictional contact with heat generation are promising next steps. Furthermore, combining VHFM with adaptive mesh refinement and higher-order spatial discretizations could enhance both accuracy and computational efficiency for large-scale, high-fidelity multiphysics simulations.

Despite its robust performance in the tested scenarios, the current formulation of VHFM has certain limitations that warrant further investigation. First, its performance in media with highly heterogeneous material properties (e.g., sharp discontinuities in thermal conductivity) requires careful examination. The method's reliance on a smoothly defined boundary normal, typically computed from a global scalar field, may be complicated by strong internal property contrasts that could disrupt the representation of the physical domain. Second, the current framework assumes boundary evolution that is continuous in topology. Its efficacy for problems involving rapid topological changes, such as fracture, fragmentation, or merging of material bodies, has not been established. The construction of a stable virtual domain surrounding dynamically changing topological features presents an open challenge. Third, while the extension to coupled problems is conceptually clear, its implementation in fully implicit, strongly coupled multiphysics systems (e.g., thermo-hydro-mechanical, where the boundary flux depends on the evolving pressure or deformation field) needs to be rigorously tested. The interaction between the virtual flux and the non-linear solution process in such coupled regimes requires additional analysis to ensure stability and consistency.

\section* {Declaration of competing interest}
\label{Declaration of competing interest}
 The authors declare that they have no known competing financial interests or personal relationships that could have appeared to influence the work reported in this paper.

\section* {Data availability}
Data will be made available on request. 

\section* {Acknowledgements}
\label{Acknowledgements} 
This work is financially supported by the National Natural Science Foundation of China (\# 52439001), Research Grants Council of Hong Kong (GRF 16208224, 16217225, CRF C7085-24G, RIF R6008-24, TRS T22-607/24N, and T22-606/23-R), and State Key Laboratory of Climate Resilience for Coastal Cities (SKLCRCC) (via Project ITC-SKLCRCC26EG01). Jidu Yu is grateful to Prof. Kenichi Soga, Dr. Bodhinanda Chandra, and Dr. Joel Given of the University of California, Berkeley, for their invaluable discussions and suggestions on this work.

\appendix
\section{Analytical solution for 1D semi-infinite rod}
\label{app:1d_analytic}
The governing equation for 1D transient heat conduction is given by,
\begin{equation}
    \frac{\partial T}{\partial t} = \alpha \frac{\partial^2 T}{\partial x^2}, \quad \text{for } x > 0, \, t > 0
    \label{eq:governing}\,,
\end{equation}
where $\alpha = \kappa / (\rho c)$ is the thermal diffusivity. The initial condition is,
\begin{equation}
    T(x, 0) = T_0, \quad \text{for } x \ge 0\,.
    \label{eq:initial}
\end{equation}

\subsection*{A.1 Constant heat flux boundary}

The boundary condition at $x=0$ is,
\begin{equation}
    -\kappa \frac{\partial T}{\partial x}\Big|_{x=0} = q_s, \quad \text{for } t > 0\,.
    \label{eq:bc_constant}
\end{equation}

This problem can be solved using the Laplace transform method. Defining the transformed temperature as $\Theta(x, s) = \mathcal{L}\{T(x, t) - T_0\}$, the governing equation becomes,
\begin{equation}
    \frac{d^2\Theta}{dx^2} - \frac{s}{\alpha}\Theta = 0\,.
\end{equation}
The solution satisfying the semi-infinite condition ($\Theta \to 0$ as $x \to \infty$) is,
\begin{equation}
    \Theta(x, s) = A(s) \exp\left(-x\sqrt{\frac{s}{\alpha}}\right)\,.
\end{equation}
Applying the transformed boundary condition from Eq.~\eqref{eq:bc_constant}, $-\kappa \frac{d\Theta}{dx}\big|_{x=0} = \frac{q_s}{s}$, yields,
\begin{equation}
    A(s) = \frac{q_s}{\kappa s^{3/2} \sqrt{\alpha}}\,.
\end{equation}
Thus,
\begin{equation}
    \Theta(x, s) = \frac{q_s}{\kappa \sqrt{\alpha}} \cdot \frac{\exp\left(-x\sqrt{s/\alpha}\right)}{s^{3/2}}\,.
\end{equation}
The inverse Laplace transform of this expression leads to the solution,
\begin{equation}
    T(x, t) = T_0 + \frac{2 q_s}{\kappa} \sqrt{\frac{\alpha t}{\pi}} \exp\left(-\frac{x^2}{4\alpha t}\right) - \frac{q_s x}{\kappa} \text{erfc}\left(\frac{x}{2\sqrt{\alpha t}}\right)\,.
    \label{eq:temp_constant_flux_inter}
\end{equation}

\subsection*{A.2 Convective heat flux boundary}
The boundary condition at $x=0$ is,
\begin{equation}
    -\kappa \frac{\partial T}{\partial x}\Big|_{x=0} = \gamma (T_a - T(0, t)), \quad \text{for } t > 0
    \label{eq:bc_convective}\,.
\end{equation}

Again, using the Laplace transform with $\Theta(x, s) = \mathcal{L}\{T(x, t) - T_0\}$, the solution in the transformed domain is,
\begin{equation}
    \Theta(x, s) = A(s) \exp\left(-x\sqrt{\frac{s}{\alpha}}\right)\,.
\end{equation}
Applying the transformed boundary condition from Eq.~\eqref{eq:bc_convective}, $-\kappa \frac{d\Theta}{dx}\big|_{x=0} = \gamma \left( \frac{T_a - T_0}{s} - \Theta(0, s) \right)$, and solving for $A(s)$ gives,
\begin{equation}
    A(s) = \frac{\gamma (T_a - T_0)/s}{\kappa\sqrt{s/\alpha} + \gamma}\,.
\end{equation}
Thus, the solution in the Laplace domain is,
\begin{equation}
    \Theta(x, s) = \frac{\gamma (T_a - T_0)}{\kappa\sqrt{\alpha}} \cdot \frac{\exp\left(-x\sqrt{s/\alpha}\right)}{\sqrt{s}\left(\sqrt{s} + \gamma / (\kappa \sqrt{\alpha}) \right)}\,.
\end{equation}
Finding the inverse Laplace transform of this expression, typically by employing a table of Laplace transforms or the convolution theorem, yields the solution,
\begin{equation}
    T(x, t) = T_0 + (T_a - T_0) \left[ \text{erfc}\left( \frac{x}{2\sqrt{\alpha t}} \right) - \exp\left( \frac{\gamma x}{\kappa} + \frac{\gamma^2 \alpha t}{\kappa^2} \right) \text{erfc}\left( \frac{x}{2\sqrt{\alpha t}} + \frac{\gamma \sqrt{\alpha t}}{\kappa} \right) \right]\,.
\end{equation}

\section{FDM algorithm for 2D circular ring}
\label{sec: FDM for ring}
For a 2D disk with a radius of $R$, the heat equation is expressed in polar coordinates $(r,\theta)$, where $r$ is the radial distance and $\theta$ is the angular coordinate. Assuming symmetry in the angular direction (so $T$ depends only on $r$ and $t$), the heat equation simplifies to,
\begin{equation}
    \frac{\partial T}{\partial t} = \alpha \left( \frac{\partial^2 T}{\partial r^2} + \frac{1}{r} \frac{\partial T}{\partial r} \right), \quad \text{for }  R_1 \leq r \leq R_2\,, \, t > 0\,.
    \label{placeholder}
\end{equation}
The initial condition is,
\begin{equation}
    T(r, 0) = T_0, \quad \text{for } R_1 \leq r \leq R_2\,.
    \label{eq:placeholder}
\end{equation}

The convective heat flux boundary conditions at the inner $(r=R_1)$ and outer boundary $(r=R_2)$ are,
\begin{align}
    &-\kappa\frac{\partial T}{\partial r}\Bigg|_{r = R_1} = \gamma\left[T(R_1,t)-T_{a}\right]\,,\\
    &-\kappa\frac{\partial T}{\partial r}\Bigg|_{r = R_2} = \gamma\left[T(R_2,t)-T_{a}\right]\,.
\end{align}

The second-order central difference and forward Euler time integration scheme are used for spatial and temporal discretizations of the equation,
\begin{equation}
T_i^{k+1} = T_i^k + \alpha \Delta t \left( \frac{T_{i+1}^k - 2T_i^k + T_{i-1}^k}{\Delta r^2} + \frac{1}{r_i} \frac{T_{i+1}^k - T_{i-1}^k}{2\Delta r} \right)\,,
\label{eq:heat_conduction}
\end{equation}
where $T_i^k$ is the temperature at radial position $r_i$ ($i=0,1,...,N$)and time step $k$, $\Delta r$ is the radial step size, $\Delta t$ is the time step size. The scheme is second-order in space and first-order in time.

At the inner and outer boundary, the second-order three-point one-Sided difference scheme is adopted,
\begin{align}
&\text{At } r = R_1 : \frac{3T_2^k - 4T_1^k + T_0^k}{2\Delta r}= 
    \gamma(T_{0}^{k}-T_a)\,,\label{eq: B5}\\
&\text{At } r = R_2 : \frac{3T_N^k - 4T_{N-1}^k + T_{N-2}^k}{2\Delta r}= 
    \gamma(T_{N}^{k}-T_a)\,,\label{eq: B6}
\end{align}
From Eqs.~\eqref{eq: B5} and \eqref{eq: B6}, the prescribed boundary temperature $T_0^k$ and $T_N^k$ can be obtained. 

For the constant heat flux boundary, it is only necessary to replace the boundary terms on the right-hand side of Eqs.~\eqref{eq: B5} and \eqref{eq: B6} with the fixed value $q_s$

\section{FDM algorithm for 3D sphere}
\label{app: 3D sphere}
For a sphere of radius $R$ with convective cooling, the 1D radial heat equation in spherical coordinates is,
\begin{equation}
    \frac{\partial T}{\partial t} = \alpha \left( \frac{\partial^2 T}{\partial r^2} + \frac{2}{r} \frac{\partial T}{\partial r} \right), \quad 0 \le r < R, \ t>0\,.
\end{equation}

The boundary conditions is 
\begin{equation}
\left.\frac{\partial T}{\partial r}\right|_{r=0} = 0, \quad
-\kappa\left.\frac{\partial T}{\partial r}\right|_{r=R} = \gamma\left[T(R,t)-T_a\right]\,,
\end{equation}
and the initial condition is,
\begin{equation}
T(r,0) = T_0\,.
\end{equation}

Similarly, using the second-order central difference and forward Euler time integration scheme for spatial and temporal discretizations of the equation,
\begin{equation}
    T_i^{k+1} = T_i^k + \alpha\Delta t\left[ \frac{T_{i+1}^k - 2T_i^k + T_{i-1}^k}{\Delta r^2} + \frac{2}{r_i} \frac{T_{i+1}^k - T_{i-1}^k}{2\Delta r} \right].
\label{eq:C4}
\end{equation}

At the inner boundary, using symmetry $T_0^k = T_2^k$,
\begin{equation}
T_1^{k+1} = T_1^k + 3\alpha\Delta t \frac{T_2^k - T_1^k}{\Delta r^2}.
\label{eq:C5}
\end{equation}

At the outer boundary, the second-order three-point one-Sided difference scheme is adopted,
\begin{equation}
-\kappa \frac{3T_N^k - 4T_{N-1}^k + T_{N-2}^k}{2\Delta r} = \gamma\left(T_N^k - T_a\right).
\label{eq:C6}
\end{equation}
From Eqs.~\eqref{eq:C4} to~\eqref{eq:C6}, the prescribed boundary temperature $T_N^k$ can be obtained.

\section{FDM algorithm for 2D square}
\label{sec: FDM for 2D square}
For 2D heat conduction in a square domain $[0, L] \times [0, L]$ with all boundaries subject to a convective heat flux boundary, the governing equation is given by,
\begin{equation}
   \frac{\partial T}{\partial t} = \alpha \left( \frac{\partial^2 T}{\partial x^2} + \frac{\partial^2 T}{\partial y^2} \right),
   \label{eq:placeholder}
\end{equation}
The boundary conditions are,
\begin{equation}
\begin{aligned}
-k \frac{\partial T}{\partial x}\Bigg|_{x = 0} &= \gamma\left[T(0,y,t)-T_{a}\right],& 
-k \frac{\partial T}{\partial x}\Bigg|_{x = L} &= \gamma\left[T(L,y,t)-T_{a}\right], \\
-k \frac{\partial T}{\partial y}\Bigg|_{y = 0} &= \gamma\left[T(x,0,t)-T_{a}\right], &
-k \frac{\partial T}{\partial y}\Bigg|_{x = L} &= \gamma\left[T(x,L,t)-T_{a}\right].
\end{aligned}
\label{eq:boundary_conditions}
\end{equation}
and the initial condition is,
\begin{equation}
    T(x, y, 0) = T_0.
    \label{eq:initial_condition}
\end{equation}

The second-order finite difference approximation for 2D heat equation is,
\begin{equation}
    T_{i,j}^{k+1} = T_{i,j}^k + \alpha \Delta t \left( \frac{T_{i+1,j}^k - 2T_{i,j}^k + T_{i-1,j}^k}{\Delta x^2} + \frac{T_{i,j+1}^k - 2T_{i,j}^k + T_{i,j-1}^k}{\Delta y^2} \right)
    \label{eq:discretized_heat}
\end{equation}
For the heat flux boundary conditions, we use a first-order finite difference approximation,
\begin{equation} \label{eq:heat_conduction}
    \begin{aligned}
        &\text{At } x = 0: & \kappa\frac{3T_{2,j}^{k} - 4T_{1,j}^{k} + T_{0,j}^{k}}{\Delta x} &= \gamma(T_{0,j}^{k}-T_a), \\
        &\text{At } x = L: & \kappa\frac{3T_{N,j}^{k} - 4T_{N-1,j}^{k} + T_{N-2,j}^{k}}{\Delta x} &= \gamma(T_{N,j}^{k}-T_a) \\
        &\text{At } y = 0: & \kappa\frac{3T_{i,2}^{k} - 4T_{i,1}^{k} + T_{i,0}^{k}}{\Delta y} &= \gamma(T_{0,i}^{k}-T_a), \\
        &\text{At } y = L: & \kappa\frac{3T_{i,N}^{k} - 4T_{i,N-1}^{k} + T_{i,N-2}^{k}}{\Delta y} &= \gamma(T_{N,i}^{k}-T_a). \\
    \end{aligned}
\end{equation}

\bibliographystyle{elsarticle-harv}
\bibliography{references}
\end{document}